\documentclass[superscriptaddress,showpacs,twocolumn,prd]{revtex4}
\usepackage{graphicx}

\usepackage{amsmath}
\newcommand{\be}{\begin{equation}}
\newcommand{\ee}{\end{equation}}
\newcommand{\ba}{\begin{eqnarray}}
\newcommand{\ea}{\end{eqnarray}}
\newcommand{\D}{{\cal D}}

\newcommand{\Tr}{\mathrm{Tr}}
\newcommand{\Str}{\mathrm{Str}}
\newcommand{\diag}{\mathrm{diag}}

\newcommand{\dsp}{\displaystyle}
\begin{document}

\onecolumngrid
\thispagestyle{empty}
\begin{flushright}
{\large 
LU  TP  06-23 \\
\large hep-lat/0606017\\[0.1cm]
\large revised August 2006}
\end{flushright}
\vskip2cm
\begin{center}
{\Large\bf
The eta mass and NNLO Three-Flavor Partially Quenched\\Chiral
Perturbation Theory}

\vfill

{\large \bf Johan Bijnens$^a$ and Niclas Danielsson$^{a,b}$}\\[1cm]
{$^a$Department of Theoretical Physics, Lund University, \\
S\"olvegatan 14A, SE - 223 62 Lund, Sweden\\[1cm]
$^b$Division of Mathematical Physics, 
Lund Institute of Technology, Lund University, \\ 
Box 118, SE - 221 00 Lund, Sweden
}

\vfill

{\large\bf Abstract}

\vskip2cm

\parbox{14cm}{\large
We show how to resum neutral propagators to all orders
in Partially Quenched Chiral Perturbation Theory.
We calculate the relevant quantities to next-to-next-to-leading order (NNLO).
Using these results we show how to extend the proposal by Sharpe
and Shoresh for determining the parameters relevant for the eta mass
from partially quenched lattice QCD calculations to NNLO.}

\vskip3cm

\end{center}
\vskip3cm
\twocolumngrid
\setcounter{page}{0}

\title{The eta mass and NNLO Three-Flavor Partially Quenched Chiral
Perturbation Theory}

\author{Johan Bijnens}
\affiliation{Department of Theoretical Physics, Lund University,\\
S\"olvegatan 14A, SE - 223 62 Lund, Sweden}
\author{Niclas Danielsson}
\affiliation{Department of Theoretical Physics, Lund University,\\
S\"olvegatan 14A, SE - 223 62 Lund, Sweden}
\affiliation{Division of Mathematical Physics, Lund Institute of 
Technology, Lund University,\\ 
Box 118, SE - 221 00 Lund, Sweden}


\begin{abstract} 
We show how to resum neutral propagators to all orders
in Partially Quenched Chiral Perturbation Theory.
We calculate the relevant quantities to next-to-next-to-leading order (NNLO).
Using these results we show how to extend the proposal by Sharpe
and Shoresh for determining the parameters relevant for the eta mass
from partially quenched lattice QCD calculations to NNLO.
\end{abstract}

\maketitle

\section{Introduction}
\label{intro}
Even though quantum chromodynamics (QCD) over time has become the generally
accepted
theory of the strong interaction, it has still proven difficult to use this
theory to derive
low-energy hadronic observables such as masses and decay constants. Lattice
QCD simulations are an alternative approach where the functional integrals are
evaluated on a discretized spacetime through numerical Monte Carlo simulations.
At present however, computational
limitations has hindered such simulations for light particles, due to the fact
that they can propagate over large distances, requiring very large
lattice sizes. Because of this, most simulations have been performed with
heavier
quark masses than those of the physical world. Typically, the quark masses
used in present lattice simulations fulfill $m_{u,d}\ge m_s/8$. In order to
get physically relevant predictions, these results then have to be extrapolated
down to the physical masses of $\sim m_s/25$.

Chiral perturbation theory ($\chi$PT) \cite{Weinberg,GL1,GL2}
is an effective field theory which provides a theoretically correct
description of the
low-energy properties of QCD. If the quark masses of the lattice simulations
are low enough so that the corresponding $\chi$PT
calculations can be considered accurate enough, this allows for a
determination of the low-energy constants (LECs) of $\chi$PT by means of a fit
to the lattice simulations. This in turn makes estimates of hadronic
low-energy observables possible. In particular, masses and decay constants
of the physical pseudoscalar mesons can be determined in this way.

It has however proven difficult to reach the chiral regime, and therefore many
lattice simulations have been performed with so-called quenched QCD. In this
theory, the effects of closed (sea) quark loops have been neglected, since such
loop effects require repeated evaluation of fermion determinants, which is
computationally extremely expensive. 
Another option is to not neglect these loop effects altogether, but instead to
introduce a separate quark mass $m_{\text{sea}}$ for the sea-quarks loops. 
Such procedures are referred
to as partial quenching (PQ), and lead into a space of unphysical theories.
This has the advantage over full QCD calculations that results with more
values of the valence quark masses can be obtained with a smaller number
of values of sea quark masses since varying the latter is computationally
more expensive.

Since unquenched QCD may be recovered from partially quenched QCD (PQQCD) in
the limit of equal sea and valence quark masses, it follows that QCD and PQQCD
are continuously connected by the variation of sea-quark masses. In
contrast, this is not true for fully quenched theories.

It has been shown that $\chi$PT can be extended to include both quenching and
partial quenching~\cite{Morel,BG1,SharpeA,BG2}. 
For partial quenching this is particularly
interesting, because it allows for determination of the physically
relevant LECs of
$\chi$PT by fits of partially quenched $\chi$PT (PQ$\chi$PT) to
partially quenched lattice simulations (PQQCD),
see e.g. the discussion in \cite{Sharpe1}. The reason for
this is that the formulation of PQ$\chi$PT is such that the dependence on the
quark masses is explicit, and thus the limit of equal sea and valence quark
masses can also be considered for PQ$\chi$PT. It follows that $\chi$PT is
recovered as a continuous limit of PQ$\chi$PT, just as QCD is from PQQCD. In
particular, the LECs of $\chi$PT, which are of physical significance, can be
obtained directly from those of PQ$\chi$PT. Clear discussions of this point
as well as calculations at next-to-leading
order (NLO) and references to earlier work
can be found in the papers of Sharpe and Shoresh
\cite{Sharpe1,Sharpe2}. There are two variants of PQ$\chi$PT possible,
with and without the supersinglet $\Phi_0$ degree of freedom.
This degree of freedom is the partially quenched analog of the singlet
$\eta'$ and is expected to be heavy compared to the other pseudoscalars for
not too large quark masses. We work in this paper with the variant without the
$\Phi_0$ \cite{Sharpe2} and with three quark flavors.

In previous papers we calculated the masses and decay constants of the charged,
or off-diagonal in flavor, pseudoscalar
mesons in PQ$\chi$PT to  next-to-next-to-leading order (NNLO) or ${\mathcal O}(p^6)$ in
the chiral counting~\cite{BDL,BL1,BL2,BDL2}. 
For these mesons and their masses the needed constants of PQ$\chi$PT to NNLO
could be determined by varying quark masses in the valence sector.
This way the structure of the extrapolation to the physical limit
of equal valence and sea quark masses and low values for the light quark masses
can be studied in detail.

Here we now turn to the neutral meson sector. The structure of the neutral 
or flavor-diagonal propagator
beyond lowest order (LO) in PQ$\chi$PT is discussed in \cite{Sharpe1,Sharpe2}.
It is shown there that the fully resummed case has double poles at the
masses related to the valence sector
as well as single poles at masses related to the
valence sector and to the neutral meson masses in the sea quark sector.
The neutral meson masses in the valence sector are the same
as the equal quark mass limit of the charged valence sector meson masses,
i.e. the mass in the valence $\overline q q$ channel is the same as in the
valence $\overline q q^\prime$ channel with $m_q=m_{q'}$. 
This is certainly not true
in full QCD.
A consequence  is that the neutral meson masses for full QCD
are not a simple limit of the neutral meson masses in the valence sector
of PQQCD but one has to study directly the meson masses in the sea-quark
sector. At NLO this can be seen by the fact the LEC $L_7^r$
cannot be determined from the masses in the valence sector but it
is known to play
a role in the eta mass \cite{GL2}. Even in the sea quark sector, $L_7^r$
can only be determined from the neutral meson masses if at least two
different sea quark masses are used, i.e. $\hat m\ne m_s$, 
since the contribution of
$L_7^r$ to the eta mass is proportional to $(\hat m-m_s)^2$.
This means that determining the chiral behavior and checking general
convergence for neutral meson masses is quite difficult in lattice
calculations. 

Sharpe and Shoresh showed in Sect. VI in \cite{Sharpe1}
that at NLO in PQ$\chi$PT a measurement of the
ratio of correlators
\be
\label{defR0}
R_0(t)\equiv\frac{\dsp \langle\pi_{ii}(t,\vec p=0)\pi_{jj}(x=0)\rangle}
{\dsp \langle\pi_{ij}(t,\vec p=0)\pi_{ji}(x=0)\rangle}
\ee
allows to determine $L_7^r$.
$\pi_{ij}$ is the source for a pseudo-scalar meson with quark flavors
$q_i\overline q_j$ and we take the valence quark masses $m_i=m_j$
but $i$ and $j$ different flavors. For large times $t$ this correlator behaves
as
\be
\label{defD}
R_0(t\to\infty) = \frac{\D t}{2 M_{ij}}\,,
\ee
with $M_{ij}$ the mass of the charged meson with quantum numbers
$q_i\overline q_j$.
The quantity $\D$ is thus a physical quantity measurable in PQQCD on the
lattice. The large $t$ behavior in (\ref{defD})
follows from the double pole structure of the
full neutral propagator~\cite{Sharpe1}. In \cite{Sharpe1,Sharpe2}
$\D$ was also studied to NLO in PQ$\chi$PT. There they showed that $L_7^r$
can be determined from $\D$, even in the case with all sea quark masses equal.
This way the LECs relevant for the eta mass can be determined from PQ$\chi$PT,
including the more detailed studies allowed by varying the valence quark mass.

$\D$ vanishes when the valence and the sea
quark masses are equal. The main part of this paper is devoted to studying
$\D$,
or more generally, the full neutral propagator, to NNLO in PQ$\chi$PT.
We therefore first show how the double pole structure from the full
neutral meson propagator follows from an all order resummation of diagrams
rather than the arguments used in \cite{Sharpe1,Sharpe2}. We have performed
this resummation for the general mass case. In the main text
we present the result only
for the simplest case. The most general case can be found
in App.~\ref{Appresum}.
Afterwards we calculate all relevant parts to NNLO
and, in particular, we
show the results for $\D$. We also discuss how $\D$ allows
to determine all needed LECs to NNLO for the eta mass.

This paper is organized in the following manner:
Section~\ref{technical_aspects} introduces the PQ$\chi$PT formalism and an
overview of the notation used for loop integrals and quark
masses. Section~\ref{resummation} contains a derivation of the resummed
neutral propagators and the double-pole coefficient ${\mathcal
  D}$. Section~\ref{D} describes how to calculate ${\mathcal D}$ to NNLO and
also contains the analytical expressions. Section~\ref{discussion} presents a
numerical analysis of the results. An extension of the fitting strategies
for extracting the various LECs from lattice QCD calculations is then found in
section~\ref{fitting_strategies}, followed by a brief discussion of our
conclusions in Section~\ref{conclusions}.

\section{$\chi$PT and  PQ$\chi$PT}
\label{technical_aspects}

In this section we give a very short overview of some aspects of $\chi$PT
and PQ$\chi$PT up to NNLO. Lectures on standard $\chi$PT can be found in
\cite{CHPTlectures}, NNLO results there are reviewed in \cite{Bijnensreview}.
A good discussion of PQ$\chi$PT can be found in \cite{Sharpe1,Sharpe2}
and the NNLO aspects are discussed in our earlier papers, but most extensively
in \cite{BDL2}. We refer to those references for more details.

In PQ$\chi$PT, a mechanism which gives different masses to sea quarks and
valence quarks is introduced in a systematic way by adding to $\chi$PT
explicit sea quarks, as well as unphysical bosonic ghost quarks. The latter
cancel exactly all effects of closed valence quarks due to their different
statistics, provided that their masses are identical to those of the valence
quarks. This results in a modification of the chiral symmetry group of
$\chi$PT. In PQ$\chi$PT, the symmetry group is given by  
\begin{equation}
\label{Gchiral}
G = SU(n_\mathrm{val}+n_\mathrm{sea} | n_\mathrm{val})_L
\times SU(n_\mathrm{val}+n_\mathrm{sea} | n_\mathrm{val})_R\,.
\end{equation}
The precise structure of $G$ is somewhat different as discussed
in~\cite{Sharpe1,Sharpe2}. This theory 
contains $n_\mathrm{val}$ valence and $n_\mathrm{val}$ ghost quarks,
as well as 
$n_\mathrm{sea}$ flavors of sea quarks. As the PQ theories include 
bosonic ghost quarks, they are not normal relativistic quantum field 
theories since they violate the spin-statistics theorem. However, under 
the assumption that the low-energy structure of such a theory can be 
described similarly to the case of normal QCD, one arrives at an 
effective low-energy theory in terms of a matrix $U$, according to
\begin{equation}
U \equiv \exp\left(i\sqrt{2}\,\Phi/\hat F\right)\,.
\end{equation}
The matrix $\Phi$ contains the Goldstone boson fields generated by
the spontaneous breakdown of the chiral symmetry group (\ref{Gchiral})
to its diagonal subgroup.
$\Phi$ has a more complicated flavor structure
than in ordinary $\chi$PT because of the different types of 
quarks present. In terms of a sub-matrix notation for the flavor 
structure
\begin{equation}
\label{SubMatrix}
q_a\bar q_b =
\left(\begin{array}{ccc}
u_a \bar u_b & u_a \bar d_b & u_a \bar s_b \\ d_a \bar u_b &
d_a \bar d_b & d_a \bar s_b \\ s_a \bar u_b & s_a \bar d_b & s_a \bar s_b    
\end{array}\right)\,,
\end{equation}
where we have used three quark flavors $u$, $d$ and $s$ in each sector, 
the matrix $\Phi$ becomes
\begin{equation}
\label{SUSY_FieldMatrix}
\Phi =
\left(\begin{array}{ccc}
\Big[\;\;q_V\bar q_V\;\;\Big] & 
\Big[\;\;q_V\bar q_S\;\;\Big] &
\Big[\;\;q_V\bar q_B\;\;\Big] \\ \\ 
\Big[\;\;q_S\bar q_V\;\;\Big] &
\Big[\;\;q_S\bar q_S\;\;\Big] & 
\Big[\;\;q_S\bar q_B\;\;\Big]\\ \\
\Big[\;\;q_B\bar q_V\;\;\Big] & 
\Big[\;\;q_B\bar q_S\;\;\Big] &
\Big[\;\;q_B\bar q_B\;\;\Big]
\end{array}\right)\,,
\end{equation}
where the labels $V,S$ and $B$ stand for valence, sea and bosonic 
quarks, respectively. The size of each sub-matrix depends on the exact 
number of quark flavors used. 

The quarks $q_V$, $q_S$ and their respective antiquarks are fermions, 
while the quarks $q_B$ and their antiquarks are bosons. Thus a 
combination of a fermionic quark and a bosonic quark yield a fermionic 
(anticommuting) field, while a combination of two fermionic or two 
bosonic quarks result in a bosonic field. Each sub-matrix in 
Eq.~(\ref{SUSY_FieldMatrix})
therefore consists of either fermionic or bosonic fields 
only. The bosonic quarks are given the 
same masses as the corresponding valence quarks in order to cancel the 
contributions from closed valence quark loops. The above formalism is 
often referred to as supersymmetric PQ$\chi$PT.

The Lagrangian structure of PQ$\chi$PT is the same as for 
$n$-flavor $\chi$PT, 
provided that the traces of 
matrix products in those Lagrangians are replaced by supertraces. The 
supertraces are defined in terms of ordinary traces by
\begin{equation}
\Str \left(\begin{array}{cc} A & B \\ C & D \end{array}\right)
=\Tr\,A - \Tr\,D\,,
\end{equation}
where $A, B, C$ and $D$ denote block matrices. For example, the block 
$B$ corresponds to the $[ q_V \bar q_B ]$ and $[ q_S \bar q_B ]$ 
sectors of the field matrix in Eq.~(\ref{SUSY_FieldMatrix}) and 
contains anticommuting fields, while the block $D$ represents the $[ 
q_B \bar q_B ]$ sector of Eq.~(\ref{SUSY_FieldMatrix}). 
A more detailed discussion about the Lagrangians and LECs can be
found in \cite{BDL2}. The expressions for the Lagrangians are in
\cite{BCE1,BCE2}. The LECs at NLO are labeled $L_i^r$ and those at NNLO
$K_i^r$. The subtraction scale dependence has been suppressed in this notation.

The version of PQ$\chi$PT considered in this paper has three flavors of 
valence quarks, three flavors of sea quarks, and 
three flavors of bosonic 
'ghost' quarks. The different quark masses are 
identified in the following calculations by the flavor indices 
$i=1,\ldots,9$, rather than by the indices $u,d,s$ and $V,S,B$ of 
Eqs.~(\ref{SubMatrix}) and~(\ref{SUSY_FieldMatrix}). The results are 
expressed in terms of the quark masses $m_q$ via the quantities 
$\chi_i=2B_0\,m_{qi}$ such that $\chi_1,\chi_2,\chi_3$, belong to the 
valence sector, $\chi_4,\chi_5,\chi_6$ to the sea sector, and 
$\chi_7,\chi_8,\chi_9$ to the ghost sector. The latter ones do not 
appear in the results since the ghost quark masses are always set equal 
to the masses of the corresponding valence quarks, such that 
$\chi_7=\chi_1, \chi_8=\chi_2$ and $\chi_9=\chi_3$. $\chi_i$ is the lowest
order meson mass squared for a charged
pseudoscalar meson with two quarks with masses
$m_i$.

For the explicit NNLO calculations in this paper
we only consider the case with all valence quark
masses equal, $\chi_1 = \chi_2 = \chi_3$, and all sea quark
masses equal, $\chi_4 = \chi_5 = \chi_6$. 
Thus only the two quark-mass parameters $\chi_1$ and $\chi_4$ appear in the
final results, but we also introduce the combinations
\be
\chi_{14} = \left(\chi_1+\chi_4\right)/2\,,\quad
R^d_{14} = \chi_1-\chi_4\,.
\ee
$\chi_{14}$ is the lowest
order meson mass squared for a pseudoscalar meson with two quarks with masses
$m_1$ and $m_4$. $R^d_{14}$ is the only one of the many combinations of
quark masses needed in the more general case~\cite{BL1,BDL2}.

The analytical expressions depend on several one- and
two-loop integrals. After carrying out the regularization and renormalization,
finite contributions remain, which are defined as
\begin{eqnarray}
\bar A(\chi) &=& -\pi_{16}\, \chi \log(\chi/\mu^2), \nonumber \\
\bar B(\chi_i,\chi_j;0) &=& -\pi_{16}\, \frac{\chi_i\log(\chi_i/\mu^2) 
- \chi_j\log(\chi_j/\mu^2)}{\chi_i - \chi_j},
\nonumber\\
\bar C(\chi,\chi,\chi;0) &=& -\pi_{16}/(2 \chi)\,.
\end{eqnarray}  
where $\mu$ denotes the renormalization scale and $\pi_{16} = 1/(16 
\pi^2)$. For the case of equal quark masses the expression for $\bar B$
reduces to
\begin{eqnarray}
\bar B(\chi,\chi;0) &=& -\pi_{16}\left(1 + \log(\chi/\mu^2) \right).
\end{eqnarray}
Some combinations of these functions are naturally generated by the
dimensional regularization procedure. We therefore define
\begin{eqnarray}
\bar A(\chi;\varepsilon) &\!=\!& \bar A(\chi)^2 / (2\pi_{16}\,\chi)
+ \pi_{16}\,\chi\,(\pi^2/12 + 1/2), \nonumber \\
\bar B(\chi,\chi;0,\varepsilon) &\!=\!& 
\bar A(\chi)\bar B(\chi,\chi;0) / (\pi_{16}\,\chi)
\!-\! \bar A(\chi)^2 / (2\pi_{16}\,\chi^2) 
\nonumber \\ && +
\pi_{16}\,(\pi^2/12 + 1/2).
\end{eqnarray}
Evaluation of the two-loop sunset diagram introduces another class of
integrals, whose
finite contributions have been denoted by $H^{F}$, $H_1^{F}$, $H_{21}^{F}$
and $H^{F'}$, $H_1^{F'}$, $H_{21}^{F'}$. The latter three functions are
derived from the former three by differentiation with respect to $p^2$ before
going on-shell. These functions all contain an integer
argument which is used to distinguish between the 8 possible configurations of
double and single poles in the integrands. A detailed discussion can be found
in~\cite{BDL2}, for PQ$\chi$PT, and in~\cite{ABT1}, 
where the single-pole version of
these integrals first appear in the context of ordinary $\chi$PT.  

\section{The Resummation}
\label{resummation}

\subsection{Introduction}

The resummation procedure is analogous to the resummation of the charged
mesons~\cite{BDL,BL1,BL2,BDL2}, except that the neutral sector has mixing.
One solution to this problem is to first diagonalize the lowest order
and then do the resummation that way. That was done for the isospin breaking
in the $\pi^0$ and $\eta$ masses in \cite{ABT2}. Doing the resummation that
way requires that the lowest order propagator matrix
in the neutral sector is invertible. 
In usual $\chi$PT this is done by working in the lowest order
eigenstate basis, i.e. one has a two by two matrix structure in the neutral
sector for three quark flavors.
\begin{figure*}
\begin{center}
\includegraphics[width=0.8\textwidth]{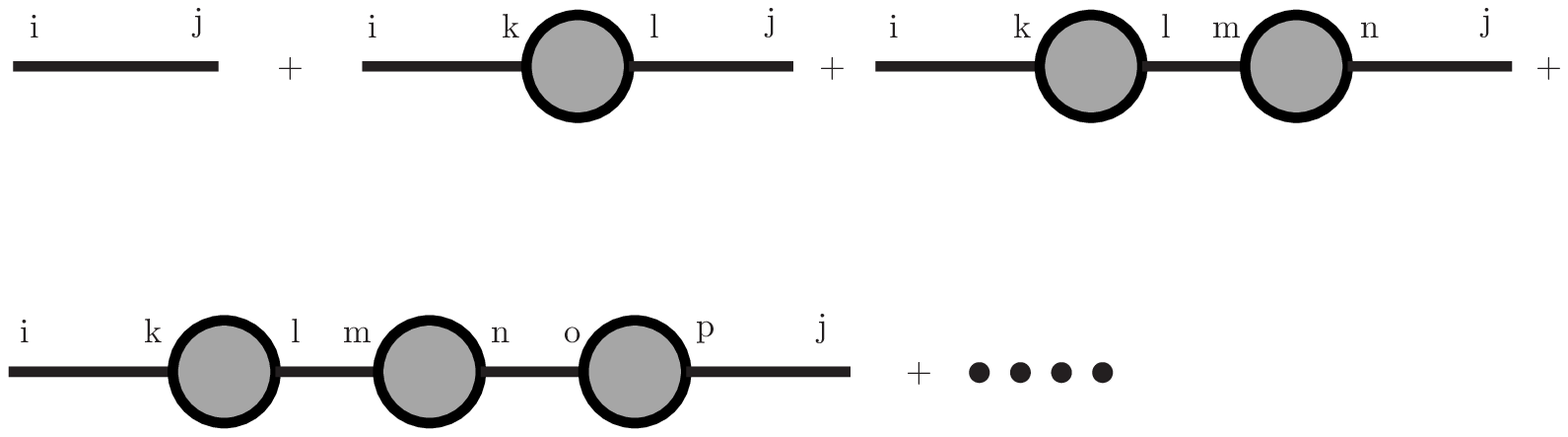}
\end{center}
\caption{\label{figTWOP}
The diagrams contributing to the
two-point function $G^n_{ij}$.
The lines are lowest order propagators.
The filled circles correspond to the sum of all one-particle-irreducible
diagrams, the self-energy.
The labels near the self-energy show the flavor index to be summed over.}
\end{figure*}

In the PQ theory, the same mechanism could be used, but this
is complicated by the
fact that one has many more states and the double poles in the lowest order
propagator as well. It is therefore easier to work in the flavor basis
rather than the lowest order eigenstate basis. After the removal
of the $\Phi_0$ degree of freedom this means that the lowest order
propagator matrix is not invertible. We thus have to perform the
resummation in a different way. Sharpe and Shoresh \cite{Sharpe1,Sharpe2}
did this by keeping the $\Phi_0$ degree of freedom and only removing it
after the resummation. Here we remove it from the start
and show that we reach the same conclusions.

We start by writing the general propagator as
\be
G_{ijkl}(x) =
\langle \Omega |T\left( \Phi_{ij}(x) \Phi_{kl}(0)\right) | \Omega \rangle\,, 
\quad \Phi_{ij}=q_i \bar q_j, 
\ee 
where $\Omega$ denotes the vacuum of the interacting theory.
Similarly, we denote the sum of all one-particle-irreducible (1PI) diagrams
with incoming quantum numbers as $\Phi_{ij}$ and outgoing as $\Phi_{kl}$,
by $-i\Sigma_{ijkl}$.
Note that this notation differs slightly from that of the previous
articles~\cite{BDL,BL1,BL2,BDL2}.
For the neutral sector we
define the $9 \times 9$ matrices in
the $(V \bar V, S \bar S, B \bar B)$ basis as the subsets of $G_{ijkl}$
and $\Sigma_{ijkl}$ with
\ba
\label{structureneutral}
G^n_{ij}(x)&\equiv& G_{iijj}(x) 
 =\langle \Omega |T\left( \Phi_{ii}(x) \Phi_{jj}(0)\right) | \Omega \rangle\,,
\nonumber\\
\Sigma_{ij}&\equiv& \Sigma_{iijj}\,.
\ea
Except for App.~\ref{Appmatrix} we will work with the Fourier transform of
these quantities.

We now look at all contributions to the full propagators.
First we recapitulate the charged propagator, $i\ne j$, case.
There we have that 
\be
\label{structurecharged}
\left.G_{ijkl}\right|_{i\ne j} = G_{ijji} \delta_{kj}\delta_{li}\,,
\ee
because of the conservation of flavor.
The lowest order propagator $\left.G^0_{ijkl}\right|_{i\ne j}$
and the self-energy $\left.\Sigma_{ijkl}\right|_{i\ne j}$ have the same 
flavor structure as Eq.~(\ref{structurecharged}).
The sum over all diagrams for the charged propagators
as depicted in Fig.~\ref{figTWOP}
becomes a diagonal sum with no internal indices to be summed over.
This resummation of the diagrams in terms of lowest order propagators
and self-energy contributions becomes a simple 
geometric series.
When summed, it gives a shift to the position of
the pole of the lowest order propagator, see e.g. ~\cite{BDL}. 
The position of this
resummed pole defines the physical masses. 

For the neutral sector the
procedure is similar, but less straightforward due to the mixing. Neither
the self-energy nor the lowest order propagator is diagonal, we thus need
to keep the nine by nine structure as defined in Eq.~(\ref{structureneutral}).
Consequently, the resummation has the structure  
\ba
G^n_{ij} &=&G^0_{ij}+G^0_{ik}(-i)\Sigma_{kl}G^0_{lj}
\nonumber\\&&
+G^0_{ik}(-i)\Sigma_{kl}G^0_{lm}(-i)\Sigma_{mn}G^0_{nj}+\cdots\,,
\label{resum_structure}
\ea
as depicted in Fig.~\ref{figTWOP}.
If $-i\Sigma G^0$ had been invertible it would have been a simple geometric
series which could be resummed via matrix inversion as done
e.g. in \cite{ABT2}. We perform below this resummation explicitly taking into
account the exact structure of the matrices $G^0$ and $-i\Sigma$.
The simplest case with all valence quark masses equal 
and all sea quark masses equal is treated here, and is referred  to as the
(1+1) case following the notation used in our earlier work.
The generalization to the more general cases can be found in
App.~\ref{Appresum}.

\subsection{Matrix Structure of $G^0$ and $-i\Sigma$.}

The lowest order neutral propagators have the general
form~\cite{BG1,Sharpe2,BDL2}
\begin{eqnarray}
G_{ij}^n (k) &=& G_{ij}^c (k)\,\delta_{ij} 
- G_{ij}^q (k) / n_\mathrm{sea}.
\label{propn}
\end{eqnarray}
The full structure can be found in App.~\ref{Appresum}
and \cite{BDL2}.

For the (1+1) mass-case we are interested in, the expression simplifies
considerably. By taking the appropriate mass limits it becomes
\be
\label{LOprop}
G^0=i\mathbf{I}\otimes\mathbf{H}-i\mathbf{K}\otimes\mathbf{D},
\ee
where $\mathbf{I}$ is the $3\times3$ unit matrix, $K$ is given by
\be
\mathbf{K}=\frac{1}{3}
\left(\begin{array}{ccc} 1 & 1 & 1 \\ 1 & 1 & 1 \\ 1 & 1 & 1 \end{array}\right)
\,, 
\quad \mathbf{K}^2=\mathbf{K}\,,
\ee
and the two matrices $\mathbf H, \mathbf D$ are given by
\be
\mathbf{H}=
\left(\begin{array}{ccc} \alpha & 0 & 0 \\ 0 & \beta & 0 \\
   0 & 0 & -\alpha\end{array}\right), 
\quad \mathbf{D}=
\left(\begin{array}{ccc} \alpha+\gamma & \alpha & \alpha+\gamma \\
  \alpha & \beta & \alpha \\ 
 \alpha+\gamma & \alpha & \alpha+\gamma\end{array}\right).
\ee
The symbols $\alpha, \beta, \gamma$ denote the different types of propagator
terms which remain after the mass limit has been taken, and are explicitly
\be
\alpha=\frac{1}{p^2-\chi_1}, \quad \beta=\frac{1}{p^2-\chi_4},
 \quad \gamma=\frac{\chi_1-\chi_4}{(p^2-\chi_1)^2}.
\label{lo_progagator_terms}
\ee

$\mathbf{I}$ and $\mathbf{K}$ give the structure inside the various sectors
as the matrix shown in Eq.~(\ref{SubMatrix}). $\mathbf{H}$ and $\mathbf{D}$
give the sector structure as in Eq.~(\ref{SUSY_FieldMatrix}).
The contributions with $\mathbf{I}$ correspond to
the connected diagrams and those with $\mathbf{K}$ to the disconnected
diagrams when thinking in terms of quark lines. Similar remarks apply to
all the other matrices discussed below.

Thus, one has a rather clean matrix structure for $G^0$, involving just the
three parameters $\alpha, \beta, \gamma$. Furthermore, since $\mathbf K$ is a
projection operator, $\mathbf{K}^2=\mathbf{K}$, 
one avoids arbitrarily high powers of $\mathbf K$.

Next, we investigate the matrix structure of the self-energy terms $-i\Sigma$.
In order to do this, we again turn the attention to the propagators $G_{ijkl}$.
For the case of one flavor of valence quarks and one flavor of sea quarks,
Sharpe and Shoresh~\cite{Sharpe1,Sharpe2} showed that several constraints on
the structure of $G_{ijkl}$ follow from the (super)symmetries inherent in the
PQ theory, and that the argument can be generalized to the more general cases.
In App.~\ref{Appmatrix}
we show this extension expicitly. From this
derivation follows that the block structure of the neutral sector, before
the $\Phi_0$ degree of freedom has been integrated out from the theory, can be
written as
\be
G=\left(\begin{array}{ccc} 
\mathbf{r+s} & \mathbf t & \mathbf r \\ \mathbf t &
    \mathbf u & \mathbf t \\ \mathbf r & \mathbf t & \mathbf{r-s} 
\end{array}\right),
\label{G_structure}
\ee
with 
\begin{eqnarray}
\mathbf s &=& s \mathbf I\,,\:\:\quad
\mathbf r = r \mathbf K\nonumber\\
\mathbf t &=& t \mathbf K\,,\quad
\mathbf u = s' \mathbf I + r' \mathbf K\,.
\label{G_blocks}
\end{eqnarray}
The symbols $s,r,t,s',r'$ denote unknown functions, whose form we are not
interested in for this discussion. We only need the structure of their
arrangement in Eq.~(\ref{G_structure}). Defining,
\begin{eqnarray}
W &=& \mathbf u^{-1}, \quad T=-\mathbf s^{-1} \mathbf t \mathbf u^{-1},\\
S &=& \mathbf s^{-1}, \quad R=\mathbf s^{-1}(\mathbf t \mathbf s^{-1}\mathbf t
\mathbf u^{-1}-\mathbf r \mathbf s^{-1}),
\end{eqnarray}
one then finds that the matrix
\be
G^{-1}=\left(\begin{array}{ccc} R+S & T & -R \\ T & W & -T \\ -R & -T & R-S 
\end{array}\right). 
\ee
is an inverse to $G$. When checking this explicitly, it is useful to note that
$\mathbf u \mathbf s^{-1} \mathbf t \mathbf u^{-1}=\mathbf t \mathbf s^{-1}$.
This matrix is a straightforward generalization of the inverse given for
the 1+1 flavor case in~\cite{Sharpe2}.    

If we now decompose $G^{-1}$ as
\be
G^{-1}=(G^0)^{-1}-i\Sigma,
\ee
where $(G^0)^{-1}$ is the inverse of the lowest order propagator matrix,
then $-i\Sigma$ corresponds to the sum of the one-particle irreducible
self-energy diagrams. 
Since $G^0$ has the same block structure as $G$, the inverse $(G^0)^{-1}$
will have the same block structure as $G^{-1}$ as well.
Thus, subtracting $(G^0)^{-1}$ from $G^{-1}$, one finds that $-i\Sigma$ has
the same block structure as well, namely
\be
\Sigma=\left(\begin{array}{ccc} R+S & T & -R \\ T & W & -T \\ -R & -T & R-S 
\end{array}\right).
\ee
where $R,S,T,W$ are different matrices from those in $G^{-1}$, 
but with the same matrix structure. The same structure should prevail after
removing the $\Phi_0$ degree of freedom~\cite{Sharpe1,Sharpe2}.
One can write $\Sigma$ in a convenient
direct product-notation as well. From the definitions of $R,S,T,W$ one can
rewrite them as
\begin{eqnarray}
R &=& \rho\mathbf K, \quad S = \sigma\mathbf I,\\
T &=& \tau\mathbf K, \quad W=w_1\mathbf I + w_2\mathbf K.
\end{eqnarray}
where $\rho,\sigma,\tau,w_1,w_2$ are as of yet unknown functions.
With this notation, $\Sigma$ can finally be written as
\be
\Sigma=\mathbf{I}\otimes\mathbf{A}+\mathbf{K}\otimes\mathbf{B},
\label{Sigma_direct_product_structure}
\ee
where
\be
\mathbf A =
\left(\begin{array}{ccc} \sigma & 0 & 0 \\ 0 & w_1 & 0 \\ 0 & 0 & -\sigma
\end{array}\right), \quad 
\mathbf B=\left(\begin{array}{ccc} \rho & \tau & -\rho \\
 \tau & w_2 & -\tau \\ -\rho & -\tau & \rho \end{array}\right).
\label{Sigma_internal_structure}
\ee

Similar arguments show that $\sigma$ is also the self-energy needed for
the off-diagonal or charged meson with both quarks having the same mass.

\subsection{The resummation}

We now perform the resummation of Eq.~(\ref{resum_structure}).
Since $\mathbf K$ is a projection operator, a typical term in the sum
looks like
\be
G^0(-i\Sigma G^0)^n=
iG^0\big(\mathbf I \otimes(\mathbf{AH})^n
+\mathbf K \otimes \{\text{all other terms}\}\big)\,.
\ee
Both $\mathbf A$ and $\mathbf H$ are diagonal, the first part thus
resums component by component as a geometric series, giving
\be
G_{\mathbf{AH}}=
\mathbf I \otimes
\left(\begin{array}{ccc} \frac{i}{p^2-\chi_1-\sigma} & 0 & 0 \\
  0 & \frac{i}{p^2-\chi_4-w_1} & 0 \\
 0 & 0 & -\frac{i}{p^2-\chi_1-\sigma}\end{array}\right)\,.
\ee
The ``$\mathbf{K}$-part'' seems less trivial. Writing out a few of the
terms in a symbolic manipulations program such as Maple, one quickly sees
a recurring structure which suggest a simple formula for the general term.
Having this formula, a proof that it holds for all orders then follows by
induction.
Denoting each $3\times 3$ block of $G$ by $(i,j), i,j=1,2,3$ we now list the
resulting expressions for each sector and resum them. 
First, the sectors $(1,1), (1,3), (3,1)$ and $(3,3)$, i.e. $r$,
all result in the same
expression for a typical term giving
\be
r_n = i\Big[ n\alpha(\rho+w_1+w_2-2\tau)
 -(n+1)\sigma(\alpha+\gamma)\Big]\sigma^{n-1}\alpha^n,
\ee
where $n=0,1,2,\dots$. For convenience, now divide $r_n$ into three parts
\be
r_n=r_n^{(1)}+r_n^{(2)}+r_n^{(3)},
\ee
with
\begin{eqnarray}
r_n^{(1)} &=& (\rho+w_1+w_2-2\tau)i\alpha^2n(\sigma\alpha)^{n-1}\nonumber\\
r_n^{(2)} &=& -i\sigma\alpha^2n(\sigma\alpha)^{n-1}\nonumber\\
        &&-i\alpha(\sigma\alpha)^n\nonumber\\
r_n^{(3)} &=&-i\gamma(n+1)(\sigma\alpha)^n\,.
\end{eqnarray}
For $r_n^{(1)}$, note that
\be
\left(\frac{1}{1-x}\right)^2 = \sum_{n=0}^{\infty}(n+1)x^n\,.
\ee
Using this and the definition of $\alpha$ from Eq.~(\ref{lo_progagator_terms}),
it follows that the $r_n^{(1)}$ resum into
\be
r^{(1)} = (\rho+w_1+w_2-2\tau)\frac{i}{(p^2-\chi_1-\sigma)^2}.
\ee
Similarly, the $r_n^{(2)}$ combine to 
\be
r^{(2)}=\frac{-i\sigma}{(p^2-\chi_1-\sigma)^2}-\frac{-i}{p^2-\chi_1-\sigma},
\ee
and, since $\gamma$ has a double pole, the $r_n^{(3)}$ sum up to
\be
r^{(3)}=\frac{-i(\chi_1-\chi_4)}{(p^2-\chi_1-\sigma)^2}\,.
\ee 
In all, we thus have the result
\begin{eqnarray}
r &=& -\frac{i}{p^2-\chi_1-\sigma}\nonumber\\&&
+\frac{i(\rho+w_1+w_2-2\tau-\sigma-(\chi_1-\chi_4))}{(p^2-\chi_1-\sigma)^2}.
\end{eqnarray}
The other sectors are not needed for this paper, since they do not contain any
double pole contributions. However, we still list the full result here for
completeness. Proceeding in the same way, we get that the neutral sector
resummed propagator becomes
\be
\label{fullG}
G=i\mathbf{I}\otimes\mathbf{H}_s-i\mathbf{K}\otimes\mathbf{D_s},
\ee
where the two matrices $\mathbf H_s, \mathbf D_s$ are given by
\be
\mathbf{H_s}=\left(\begin{array}{ccc} \bar \alpha & 0 & 0 \\ 0 & \bar \beta & 0
    \\ 0 & 0 & -\bar \alpha\end{array}\right), \quad
\mathbf{D_s}=\left(\begin{array}{ccc} \bar \alpha+\bar \gamma & \bar \alpha &
    \bar \alpha+\bar \gamma \\ \bar \alpha & \bar \beta & \bar \alpha \\ \bar
  \alpha+\bar \gamma & \bar \alpha & \bar \alpha+\bar \gamma\end{array}\right).
\ee
The symbols $\bar \alpha, \bar \beta, \bar \gamma$ denote the various resummed
propagator terms, given by
\begin{eqnarray}
\bar \alpha &=& \frac{1}{p^2-\chi_1-\sigma}, \quad \bar
\beta=\frac{1}{p^2-\chi_4-w_1},\nonumber\\
\bar \gamma &=& 
 \frac{\chi_1-\chi_4+\sigma+2\tau-\rho-w_1-w_2}{(p^2-\chi_1-\sigma)^2}.
\end{eqnarray}
We  see that the full propagator has exactly the same
structure as the lowest order propagator in this simplest mass case.
The resummed propagator matrix involves the three parameters $\bar \alpha, \bar
\beta, \bar \gamma$,
which are the direct counterparts to the lowest order parameters $\alpha,
\beta, \gamma$ of $G^0$.

\subsection{$\D$ from $G$}

We now expand the double pole around the position of the pole. The position is
the same as the charged mass squared since $\sigma$ is also the self-energy for
that case. The position of the zero of $p^2-\chi_1-\sigma$ we thus refer to
as $M_{ch}^2$. It is the physical mass of the charged mesons (in
the degenerate mass-case, there is only one such mass).
We expand around the double pole term
and get
\be
G^n_{ij} = 
\frac{-i{\cal Z}{\cal D}}{(p^2-M^2_{ch})^2}+\cdots\,.
\label{D_def}
\ee
The renormalization factor $\cal Z$ is defined as
\be
\frac{1}{\cal Z}= 1-\left.\frac{\partial
    \sigma(p^2,\chi_i)}{\partial p^2}\right\vert_{p^2=M_{ch}^2}.
\ee
Here and below we have indicated the dependence on the momentum and the
$\chi_i$ of the various quantities. They also depend on the LECs
of PQ$\chi$PT.
The definition of $\D$ in (\ref{D_def})
coincides with the one given in Eq.~(\ref{defD}).

Eqs.~(\ref{fullG}) and (\ref{D_def}) then imply that $\cal D$ is given by
\be
{\cal D}=\frac{1}{3}{\cal Z}\Big(\chi_1-\chi_4+\sigma+2\tau-\rho-w_1-w_2\Big).
\label{D_symm}
\ee
Note that the factor $1/3$ from $\mathbf K$ has been included here as well.

\section{Analytical results for ${\cal D}$}
\label{D}

Equation~(\ref{D_symm}) is written in terms of the blocks of
$\Sigma_{ij}$, but
for the explicit calculations we want to have the expression in terms of the
self-energies $\Sigma_{ij}$ directly, since this is what we know how to
calculate. In
the degenerate case we are studying in this paper, there are several
equivalent ways
to obtain the quantities needed since $\sigma,\rho,\tau,w_1$ and
$w_2$ appear in many different elements of the full self-energy matrix.
We chose to use
\begin{eqnarray}
\sigma&=& \Sigma_{11}-\Sigma_{12}\,,\quad
\rho = 3\Sigma_{12}\nonumber\\
w_1&=&\Sigma_{44}-\Sigma_{45}\,,\quad
w_2 = 3\Sigma_{45}\,,\quad
\tau=3\Sigma_{14}.
\end{eqnarray} 
which follows immediately from
Eqs.~(\ref{Sigma_direct_product_structure}) and
(\ref{Sigma_internal_structure}). 
Then we get
\be
{\cal D}=\frac{{\cal Z}}{3}\Big(\chi_1-\chi_4+\Sigma_{11}
   -4\Sigma_{12}-\Sigma_{44}-2\Sigma_{45}+6\Sigma_{14}\Big).
\label{DD}
\ee

\subsection{${\cal D}$ at NNLO}

For the NNLO calculation of ${\cal D}$, we have to determine the expression
for~(\ref{DD}) up to ${\mathcal O} (p^6)$. This means that all quantities appearing
in Eq.~(\ref{D_symm}) have to be expanded up to the appropriate order.

The first terms are of the form 
$(\chi_1-\chi_4){\mathcal Z}$. 
Here $\chi_1-\chi_4$ is of $\mathcal{O}(p^2)$, so we have to
determine ${\mathcal Z}$ up to $\mathcal{O}(p^4)$. From the definition of
${\mathcal Z}$ and 
$\sigma'(M^2_{ch})=\partial\sigma(p^2,\chi_i)/\partial p^2|_{p^2=M_{ch}^2}$,
we get
\be
{\cal Z}=\frac{1}{1-\sigma'(M^2_{ch})}
=1+\sigma'(M^2_{ch})+\sigma'(M^2_{ch})^2+\dots
\ee
The expressions for $\sigma'(M^2_{ch})$ can be written as a string of terms
which denote the 1PI diagrams of progressively higher order.
\be
\sigma'(M_{{ch}}^2,\chi_i) = 
{\sigma'}^{(4)}(M_{{ch}}^2,\chi_i) 
+{\sigma'}^{(6)}(M_{{ch}}^2,\chi_i) \:+\: \cdots
\label{sigmadiagr}
\ee
where ${\sigma'}^{(4)}$ contains derivatives of the self-energy diagrams of 
${\mathcal O}(p^4)$, and ${\sigma'}^{(6)}$ derivatives of those of
${\mathcal O}(p^6)$. Note that differentiation lowers the order of the
diagram contributions by two, and therefore these terms are of
${\mathcal O}(p^2)$ and ${\mathcal O}(p^4)$ respectively.
It is sufficient to use the lowest order mass instead of $M_{\mathrm{ch}}^2$ 
in ${\sigma'}^{(6)}$ since the differentiated diagrams in that term are
already of 
${\mathcal O}(p^4)$. However, in the case of ${\sigma'}^{(4)}$ the argument 
$M_{{ch}}^2$ should be expanded in a Taylor series around 
$M_0^2=\chi_1$, since the diagrams in ${\sigma'}^{(4)}$ are of
$\mathcal{O}(p^2)$. If $M_{\mathrm{ch}}^2$ is formally written as
\begin{eqnarray}
M_{{ch}}^2 &=& M_0^2 + M_4^2 + M_6^2 +\cdots\,,
\end{eqnarray}
then an expansion of ${\sigma'}^{(4)}$ up to ${\mathcal O}(p^4)$ 
gives
\begin{eqnarray}
{\sigma'}^{(4)}(M_{{ch}}^2,\chi_i) &=& {\sigma'}^{(4)}(M_0^2,\chi_i) 
\:+\: M_4^2 \left.\frac{\partial {\sigma'}^{(4)}(p^2,\chi_i)}{\partial p^2}
\right\vert_{M_0^2} \hspace{-.5cm} \nonumber \\
&+& \mathcal{O}(p^6),
\label{taylorsigma}
\end{eqnarray}
where $M_4^2$ represents the NLO correction to the lowest order charged
meson mass, which is given by $\sigma^{(4)}(M_0^2,\chi_i)$. In practice,
however, this Taylor expanded term is not needed, because $\sigma^{(4)}$
contains no higher powers of $p^2$, and thus the derivative term
of~(\ref{taylorsigma}) is identically zero.

Collecting the relevant terms up to ${\mathcal O}(p^6)$ gives
\be
{\mathcal Z} \!=\!
 1 + {\sigma'}^{(4)}(M_0^2,\chi_i)
+
 {\sigma'}^{(6)}(M_0^2,\chi_i)
+ \left({\sigma'}^{(4)}(M_0^2,\chi_i)\right)^2.
\ee

Next, the form of ${\mathcal Z} \Sigma_{ij}$ has to 
be determined up to ${\mathcal O}(p^6)$. In order to do this, we determine
$\Sigma_{ij}$ up to ${\mathcal O}(p^6)$, multiply the expressions for
${\mathcal Z}$ and $\Sigma_{ij}$ together and collect the resulting terms up
to ${\mathcal O}(p^6)$.
As for the $\sigma$ terms we write the self-energies $\Sigma_{ij}$ as a
string of terms of progressively 
higher order:
\be
\Sigma_{ij}(M_{{ch}}^2,\chi_i) = 
\Sigma_{ij}^{(4)}(M_{{ch}}^2,\chi_i)
+\Sigma_{ij}^{(6)}(M_{{ch}}^2,\chi_i) + \cdots\,.
\ee
For $\Sigma_{ij}^{(4)}$ the argument 
$M_{{ch}}^2$ must be expanded in a Taylor series around 
$M_0^2=\chi_1$, since the diagrams in $\Sigma_{ij}^{(4)}$ are of
${\mathcal O}(p^4)$. 
An expansion of $\Sigma_{ij}^{(4)}$ up to ${\mathcal O}(p^6)$ 
gives
\begin{eqnarray}
\Sigma_{ij}^{(4)}(M_{\mathrm{ch}}^2,\chi_i) &=& \Sigma_{ij}^{(4)}(M_0^2,\chi_i) 
\:+\: M_4^2 \left.\frac{\partial \Sigma_{ij}^{(4)}(p^2,\chi_i)}{\partial p^2}
\right\vert_{M_0^2} \hspace{-.5cm} \nonumber \\
&+& {\mathcal O}(p^8),
\end{eqnarray}
where $M_4^2$ again represents the NLO correction to the lowest order charged
meson mass, given by $\sigma^{(4)}(M_0^2,\chi_i)$. 

Thus, up to ${\mathcal O}(p^6)$, the $\Sigma_{ij}$'s are given by
\ba
\Sigma_{ij} &=& \Sigma_{ij}^{(4)}(M_0^2,\chi_i) 
+ \Sigma_{ij}^{(6)}(M_0^2,\chi_i)
\nonumber \\ 
&+& \sigma^{(4)}(M_0^2,\chi_i) 
\left.\frac{\partial \Sigma_{ij}^{(4)}(p^2,\chi_i)}{\partial p^2}
\right\vert_{M_0^2}
 \:+\:{\mathcal O}(p^8),
\label{Sigmaeq}
\ea
where the last two terms are of ${\mathcal O}(p^6)$ and represent the 
NNLO contributions. For convenience, we will use the notation
${\Sigma'} _{ij}^{(4)}$
for the derivative term, assuming it understood that the derivative is as in
Eq.~(\ref{Sigmaeq}). 

Having this result, we now get the final expression for
${\cal  Z}\Sigma_{ij}$ up to ${\mathcal O}(p^6)$ to be
\ba
{\cal Z}\Sigma_{ij} &=& \Sigma_{ij}^{(4)}(M_0^2,\chi_i)
+ \Sigma_{ij}^{(6)}(M_0^2,\chi_i)\nonumber\\
&+& \sigma^{(4)}(M_0^2,\chi_i){\Sigma
  '}_{ij}^{(4)}(M_0^2,\chi_i)\nonumber\\
&+& {\sigma'}^{(4)}(M_0^2,\chi_i)\Sigma_{ij}^{(4)}(M_0^2,\chi_i).
\ea

Putting everything together, the analytical expression for 
${\cal D}$ up to ${\mathcal O}(p^6)$,
or NNLO, is given by
\ba
{\mathcal D}^{(6)} &\!=\!& \frac{1}{3}\Bigg\{ 
(\chi_1-\chi_4)\left[ 1 + {\sigma'}^{(4)}{\sigma'}^{(6)}
+\left({\sigma'}^{(4)}\right)^2\right]
\nonumber\\&&
+\Sigma_{11}^{(4)}+\sigma^{(4)}{\Sigma
  '}_{11}^{(4)}+{\sigma'}^{(4)}\Sigma_{11}^{(4)}+\Sigma_{11}^{(6)}
\nonumber\\&&
-4\left(\Sigma_{12}^{(4)}+\sigma^{(4)}{\Sigma
  '}_{12}^{(4)}+{\sigma'}^{(4)}\Sigma_{12}^{(4)}+\Sigma_{12}^{(6)}\right)
\nonumber\\&&
+6\left(\Sigma_{14}^{(4)}+\sigma^{(4)}{\Sigma
  '}_{14}^{(4)}+{\sigma'}^{(4)}\Sigma_{14}^{(4)}+\Sigma_{14}^{(6)}\right)
\nonumber\\&&
-\;\;\left(\Sigma_{44}^{(4)}+\sigma^{(4)}{\Sigma
  '}_{44}^{(4)}+{\sigma'}^{(4)}\Sigma_{44}^{(4)}+\Sigma_{44}^{(6)}\right)
\nonumber\\&&
-2\left( \Sigma_{45}^{(4)}+\sigma^{(4)}{\Sigma
  '}_{45}^{(4)}+{\sigma'}^{(4)}\Sigma_{45}^{(4)}+\Sigma_{45}^{(6)}\right)
\!\!\Bigg\}.
\ea
where the arguments of the functions have been suppressed.

\subsection{Expression for ${\cal D}$}

The analytical expression for ${\mathcal D}$ is given to NNLO in the form
\be
\label{defDNNLO}
{\mathcal D}= {\mathcal D}^{(2)}+\frac{{\mathcal D}^{(4)}}{F^2_0}
+\frac{{\mathcal D}_{\text{ct}}^{(6)}+{\mathcal D}_{\text{loop}}^{(6)}}{F^4_0}
+{\mathcal O}(p^8)\,,
\ee  
where the ${\mathcal O}(p^4)$ and ${\mathcal O}(p^6)$ contributions have been separated. The 
NNLO contribution ${\mathcal D}^{(6)}$ has been further split into the 
contributions from the chiral loops and from the ${\mathcal O}(p^6)$ 
counterterms or LECs.

It should be 
noted that we have chosen to give the results to the various orders in 
terms of the three flavor lowest order decay constant $F_0$
and in terms of the 
lowest order meson masses, since these are the fundamental inputs in 
PQ$\chi$PT. The situation is different in standard $\chi$PT, where the 
main objective is comparison with experiment, in which case the 
formulas are most often rewritten in terms of the physical decay 
constants and masses. We have changed here the general $\hat F$ to $F_0$
as is appropriate for the three flavor case.

The lowest order contribution to ${\mathcal D}$ is just the difference between
the valence sector quark mass and the sea sector quark mass, i.e 
\be
   {\cal{D}}^{(2)} =\frac{1}{3} R^d_{14}.
\ee
Here one sees explicitly the fact that this is a quantity relevant only in the
partially quenched theory, since the lowest order contribution vanishes when
one takes the limit $\chi_4\rightarrow \chi_1$ of unquenched $\chi$PT. We have
checked that this also holds for the ${\cal O}(p^4)$ and ${\cal O}(p^6)$
contributions. 

The combined NLO contribution to ${\mathcal D}$ (loops and counterterms), is
\ba
\label{resultp4}
{\cal{D}}^{(4)} &=&\frac{1}{3}\big\{
       -24L_4^r R^d_{14}\chi_4 
       -16L_5^r R^d_{14}\chi_1 
       +48L_6^r R^d_{14}\chi_4 
\nonumber\\&&
       -48L_7^r (R^d_{14})^2 
       +32L_8^r R^d_{14}\chi_{14} 
       -10/3 \bar A(\chi_1)\chi_1 
\nonumber\\&&
       +6 \bar A(\chi_{14})\chi_{14} 
       -8/3 \bar A(\chi_4) \chi_4 
\nonumber\\&&
       -1/3 \bar B(\chi_1,\chi_1;0) R^d_{14}\chi_1\big\}. 
\ea
and is in agreement with Refs.~\cite{Sharpe1,Sharpe2}.
The NNLO contribution from the ${\mathcal O}(p^6)$ counterterms is given by
\ba
\label{resultp6Ki}
{\cal{D}}_{\text{ct}}^{(6)} &=&\frac{1}{3}\big\{
       -96 K_{17}^r \chi_1^2R^d_{14}
       -192 K_{18}^r \chi_1\chi_4R^d_{14} 
\nonumber\\&&
       + K_{19}^r  [ 16\chi_1(R^d_{14})^2 - 48\chi_1^2R^d_{14} ] 
       -96 K_{20}^r  \chi_1\chi_4R^d_{14} 
\nonumber\\&& 
      -48 K_{21}^r \chi_4^2R^d_{14}
       -144 K_{22}^r \chi_4^2R^d_{14}
\nonumber\\&&
       + K_{23}^r  [ 16\chi_1(R^d_{14})^2 - 48\chi_1^2R^d_{14} ] 
       +48 K_{24}^r \chi_1(R^d_{14})^2 
\nonumber\\&&
       +48 K_{25}^r  [\chi_1^3 - \chi_4^3 ] 
       + K_{26}^r  [ 96\chi_1\chi_4R^d_{14} + 144\chi_4^2R^d_{14} ] 
\nonumber\\&&
       +432 K_{27}^r \chi_4^2R^d_{14} 
       +32 K_{39}^r  [\chi_1^3 - \chi_4^3 ] 
\nonumber\\&&
       +192 K_{40}^r \chi_{14}\chi_4R^d_{14} 
       -192 K_{41}^r \chi_{14}(R^d_{14})^2 
\nonumber\\&&
       -288 K_{42}^r \chi_4(R^d_{14})^2\big\}. 
\ea

The NNLO loop contribution is considerably longer. Therefore we have made the
\texttt{FORM}~\cite{FORM} output with
all the analytical expressions available for download from the web
site~\cite{website}. The NNLO loop result is
\begin{widetext}
\ba
\label{resultp6loops}
   {\cal{D}}_{\text{loop}}^{(6)} &=&\frac{1}{3}\big\{
       \pi_{16}L_0^r  [ 49/3\chi_1\chi_4R^d_{14} - 2\chi_1^2R^d_{14} + 3\chi_4^2R^d_{14} ] 
       +12 \pi_{16}L_1^r \chi_1^2R^d_{14} 
       + \pi_{16}L_2^r  [ 6\chi_1^2R^d_{14} + 16\chi_4^2R^d_{14} ] 
\nonumber\\&&
       + \pi_{16}L_3^r  [ 71/6\chi_1\chi_4R^d_{14} - 8\chi_1^2R^d_{14} + 3/2\chi_4^2R^d_{14} ] 
       + \pi_{16}^2  [ 21/32\chi_1\chi_4R^d_{14} + 73/64\chi_4^2R^d_{14} ] 
       +768 L_4^rL_5^r \chi_1\chi_4R^d_{14} 
\nonumber\\&&
       -1152 L_4^rL_6^r \chi_4^2R^d_{14} 
       +1152 L_4^rL_7^r \chi_4(R^d_{14})^2 
       -768 L_4^rL_8^r \chi_{14}\chi_4R^d_{14} 
       +576 L_4^{r2} \chi_4^2R^d_{14} 
       -768 L_5^rL_6^r \chi_1\chi_4R^d_{14} 
\nonumber\\&&
       +384 L_5^rL_7^r \chi_1(R^d_{14})^2 
       + L_5^rL_8^r  [ 128\chi_1(R^d_{14})^2 - 384\chi_1^2R^d_{14} ] 
       +192 L_5^{r2} \chi_1^2R^d_{14} 
       + \bar A(\chi_1)\pi_{16}  [ 9/4\chi_1\chi_4 + 3/4\chi_1^2 ] 
\nonumber\\&&
       + \bar A(\chi_1)L_0^r  [ 32\chi_1\chi_4 - 80\chi_1^2 - 8\chi_4^2 ] 
       +16 \bar A(\chi_1)L_1^r \chi_1R^d_{14} 
       +40 \bar A(\chi_1)L_2^r \chi_1R^d_{14} 
\nonumber\\&&
       + \bar A(\chi_1)L_3^r  [ 32\chi_1\chi_4 - 44\chi_1^2 - 8\chi_4^2 ] 
       +160 \bar A(\chi_1)L_4^r \chi_1\chi_4 
       + \bar A(\chi_1)L_5^r  [  - 32/3\chi_1\chi_4 + 208/3\chi_1^2 ]
\nonumber\\&&
       + \bar A(\chi_1)L_6^r  [  - 224\chi_1\chi_4 + 64\chi_1^2 ] 
       +64 \bar A(\chi_1)L_7^r \chi_1R^d_{14} 
       -352/3 \bar A(\chi_1)L_8^r \chi_1^2 
       +14/9 \bar A(\chi_1)^2 \chi_1 
\nonumber\\&&
       +7/2 \bar A(\chi_1)\bar A(\chi_{14}) \chi_1 
       + \bar A(\chi_1)\bar B(\chi_1,\chi_1;0)  [  - 10/9\chi_1\chi_4 + 10/3\chi_1^2 ] 
       -2 \bar A(\chi_1)\bar B(\chi_{14},\chi_{14};0) \chi_{14}^2 
\nonumber\\&&
       +2/9 \bar A(\chi_1)\bar C(\chi_1,\chi_1,\chi_1;0) \chi_1^2R^d_{14} 
       + \bar A(\chi_{14})\pi_{16}  [  - 3/4\chi_1R^d_{14} - 3\chi_4^2 ] 
       + \bar A(\chi_{14})L_0^r  [ 144\chi_1\chi_{14} - 24\chi_{14}\chi_4 ] 
\nonumber\\&&
       + \bar A(\chi_{14})L_3^r  [ 144\chi_1\chi_{14} - 60\chi_{14}\chi_4 ] 
       -288 \bar A(\chi_{14})L_4^r \chi_{14}\chi_4 
       -144 \bar A(\chi_{14})L_5^r \chi_1\chi_{14} 
       +288 \bar A(\chi_{14})L_6^r \chi_{14}\chi_4 
\nonumber
\ea
\end{widetext}
\begin{widetext}
\ba
\phantom{{\cal{D}}_{\text{loop}}^{(6)}} &\phantom{=}&
       -288 \bar A(\chi_{14})L_7^r \chi_{14}R^d_{14} 
       + \bar A(\chi_{14})L_8^r  [ 192\chi_1\chi_{14} + 96\chi_{14}\chi_4 ] 
       -27/2 \bar A(\chi_{14})^2 \chi_{14} 
       +8 \bar A(\chi_{14})\bar A(\chi_4) \chi_4 
\nonumber\\&&
       -2 \bar A(\chi_{14})\bar B(\chi_1,\chi_1;0) \chi_1\chi_{14} 
       -64 \bar A(\chi_4)L_0^r \chi_1\chi_4 
       +128 \bar A(\chi_4)L_1^r \chi_4R^d_{14} 
       +32 \bar A(\chi_4)L_2^r \chi_4R^d_{14} 
       -64 \bar A(\chi_4)L_3^r \chi_1\chi_4 
\nonumber\\&&
       + \bar A(\chi_4)L_4^r  [  - 128\chi_1\chi_4 + 256\chi_4^2 ] 
       +256/3 \bar A(\chi_4)L_5^r \chi_{14}\chi_4
       + \bar A(\chi_4)L_6^r  [ 128\chi_1\chi_4 - 256\chi_4^2 ] 
       +256 \bar A(\chi_4)L_7^r \chi_4R^d_{14}
\nonumber\\&&
       -512/3 \bar A(\chi_4)L_8^r \chi_4^2 
       +4/9 \bar A(\chi_4)^2 \chi_4 
       +8/9 \bar A(\chi_4)\bar B(\chi_1,\chi_1;0) \chi_1\chi_4 
       +8/9 \bar A(\chi_4)\bar B(\chi_4,\chi_4;0) \chi_4^2 
\nonumber\\&&
       + \bar A(\chi_1,\epsilon)\pi_{16}  [ 7/36\chi_1\chi_4 + 11/18\chi_1^2 + 1/4\chi_4^2 ] 
       + \bar A(\chi_{14},\epsilon)\pi_{16}  [  - 19/2\chi_1\chi_4 - 5/2\chi_1^2 + 9/2\chi_4^2 ] 
\nonumber\\&&
       +58/9 \bar A(\chi_4,\epsilon)\pi_{16} \chi_4^2 
       + \bar B(\chi_1,\chi_1;0)L_0^r  [ 8\chi_1\chi_4R^d_{14} - 16\chi_1^2R^d_{14} ] 
       +8 \bar B(\chi_1,\chi_1;0)L_1^r \chi_1^2R^d_{14} 
\nonumber\\&&
       +20 \bar B(\chi_1,\chi_1;0)L_2^r \chi_1^2R^d_{14} 
       + \bar B(\chi_1,\chi_1;0)L_3^r  [ 8\chi_1\chi_4R^d_{14} - 16\chi_1^2R^d_{14} ] 
       + \bar B(\chi_1,\chi_1;0)L_4^r  [  - 24\chi_1\chi_4^2 + 104\chi_1^2\chi_4 ] 
\nonumber\\&&
       + \bar B(\chi_1,\chi_1;0)L_5^r  [ 16/3\chi_1\chi_4^2 - 80/3\chi_1^2\chi_4 + 48\chi_1^3 ] 
       + \bar B(\chi_1,\chi_1;0)L_6^r  [ 32\chi_1(R^d_{14})^2 - 160\chi_1^2\chi_4 ] 
\nonumber\\&&
       +48 \bar B(\chi_1,\chi_1;0)L_7^r \chi_1(R^d_{14})^2 
       + \bar B(\chi_1,\chi_1;0)L_8^r  [ 16/3\chi_1\chi_4^2 + 64/3\chi_1^2\chi_4 - 80\chi_1^3 ] 
\nonumber\\&&
       + \bar B(\chi_1,\chi_1;0)^2  [  - 1/18\chi_1\chi_4R^d_{14} + 23/18\chi_1^2R^d_{14} ] 
       +2/9 \bar B(\chi_1,\chi_1;0)\bar C(\chi_1,\chi_1,\chi_1;0) \chi_1^2(R^d_{14})^2 
\nonumber\\&&
       -144 \bar B(\chi_{14},\chi_{14};0)L_4^r \chi_{14}^2\chi_4 
       -48 \bar B(\chi_{14},\chi_{14};0)L_5^r \chi_{14}^3 
       +288 \bar B(\chi_{14},\chi_{14};0)L_6^r \chi_{14}^2\chi_4 
       +96 \bar B(\chi_{14},\chi_{14};0)L_8^r \chi_{14}^3 
\nonumber\\&&
       +64 \bar B(\chi_4,\chi_4;0)L_4^r \chi_4^3 
       +64/3 \bar B(\chi_4,\chi_4;0)L_5^r \chi_4^3 
       -128 \bar B(\chi_4,\chi_4;0)L_6^r \chi_4^3 
       -128/3 \bar B(\chi_4,\chi_4;0)L_8^r \chi_4^3 
\nonumber\\&&
       + \bar B(\chi_1,\chi_1;0,\epsilon)\pi_{16}  [  - 1/4\chi_1\chi_4R^d_{14} - 2/3\chi_1^2R^d_{14} ] 
       +16 \bar C(\chi_1,\chi_1,\chi_1;0)L_4^r \chi_1^2\chi_4R^d_{14} 
\nonumber\\&&
       +16/3 \bar C(\chi_1,\chi_1,\chi_1;0)L_5^r \chi_1^3R^d_{14} 
       -32 \bar C(\chi_1,\chi_1,\chi_1;0)L_6^r \chi_1^2\chi_4R^d_{14} 
       -32/3 \bar C(\chi_1,\chi_1,\chi_1;0)L_8^r \chi_1^3R^d_{14} 
\nonumber\\&&
       + H^F(1,\chi_1,\chi_1,\chi_1,\chi_1)  [  - 4/27\chi_1\chi_4 + 44/27\chi_1^2 ] 
       +5/9 H^{F'}(1,\chi_1,\chi_1,\chi_1,\chi_1) \chi_1^2R^d_{14} 
\nonumber\\&&
       + H^F(1,\chi_1,\chi_{14},\chi_{14},\chi_1)  [  - 10\chi_1^2 + 1/4\chi_4R^d_{14} ] 
       + H^{F'}(1,\chi_1,\chi_{14},\chi_{14},\chi_1)  [ 1/4\chi_1\chi_4R^d_{14} - \chi_1^2R^d_{14} ] 
\nonumber\\&&
       -2 H^F(1,\chi_{14},\chi_{14},\chi_4,\chi_1) \chi_4^2 
       +2 H^{F'}(1,\chi_{14},\chi_{14},\chi_4,\chi_1) \chi_1\chi_4R^d_{14} 
       -40/27 H^F(1,\chi_4,\chi_4,\chi_4,\chi_1) \chi_4^2 
\nonumber\\&&
       +5/3 H^F(2,\chi_1,\chi_1,\chi_1,\chi_1) \chi_1^2R^d_{14} 
       +4/9 H^{F'}(2,\chi_1,\chi_1,\chi_1,\chi_1) \chi_1^2(R^d_{14})^2 
\nonumber\\&&
       + H^F(2,\chi_1,\chi_{14},\chi_{14},\chi_1)  [ 1/4\chi_1\chi_4R^d_{14} - \chi_1^2R^d_{14} ] 
       -5/4 H^{F'}(2,\chi_1,\chi_{14},\chi_{14},\chi_1) \chi_1^2(R^d_{14})^2 
\nonumber\\&&
       +4/9 H^F(5,\chi_1,\chi_1,\chi_1,\chi_1) \chi_1^2(R^d_{14})^2 
       +2/9 H^{F'}(5,\chi_1,\chi_1,\chi_1,\chi_1) \chi_1^2(R^d_{14})^3 
\nonumber\\&&
       +2/27 H^F(8,\chi_1,\chi_1,\chi_1,\chi_1) \chi_1^2(R^d_{14})^3 
       + H_1^F(1,\chi_1,\chi_{14},\chi_{14},\chi_1)  [ 2\chi_1\chi_4 + 18\chi_1^2 ] 
\nonumber\\&&
       +16 H_1^F(1,\chi_4,\chi_{14},\chi_{14},\chi_1) \chi_1\chi_4 
       +2 H_1^F(2,\chi_1,\chi_{14},\chi_{14},\chi_1) \chi_1(R^d_{14})^2 
       +2 H_1^{F'}(2,\chi_1,\chi_{14},\chi_{14},\chi_1) \chi_1^2(R^d_{14})^2 
\nonumber\\&&
       + H_{21}^F(1,\chi_1,\chi_{14},\chi_{14},\chi_1)  [  - 3/2\chi_1\chi_4 - 6\chi_1^2 ] 
       +3/4 H_{21}^{F'}(1,\chi_1,\chi_{14},\chi_{14},\chi_1) \chi_1^2R^d_{14} 
\nonumber\\&&
       +27/2 H_{21}^F(1,\chi_{14},\chi_1,\chi_{14},\chi_1) \chi_1^2 
       + H_{21}^F(1,\chi_4,\chi_{14},\chi_{14},\chi_1)  [  - 12\chi_1\chi_4 + 6\chi_1^2 ] 
       +6 H_{21}^{F'}(1,\chi_4,\chi_{14},\chi_{14},\chi_1) \chi_1^2R^d_{14} 
\nonumber\\&&
       + H_{21}^F(2,\chi_1,\chi_{14},\chi_{14},\chi_1)  [ 3/2\chi_1\chi_4R^d_{14} - 3/4\chi_1^2R^d_{14} ] 
  -3/4 H_{21}^{F'}(2,\chi_1,\chi_{14},\chi_{14},\chi_1) \chi_1^2(R^d_{14})^2
\big\}.
\ea
\end{widetext}

\section{Numerical Results and Discussion}
\label{discussion}

\subsection{Checks on the Calculation}
\label{checks}

As for the previous articles on NNLO calculations in PQ$\chi$PT, the
calculations are rather lengthy, and require careful checking procedures to
avoid the possibility for mistakes showing up in the end result. The complete
calculations have therefore been performed in two complete independent
versions, which have been checked against each other for agreement. The
numerical results have been treated likewise, and the programs were written in
different programming languages (Fortran and C++).
Furthermore, the divergence structure agrees with general calculation
of \cite{BCE2}
and the cancellation of nonlocal divergences happens as required.
The end results are expected to vanish for $\chi_4\rightarrow
\chi_1$, since the quantity ${\mathcal D}$ is a quantity relevant for the PQ
theories only. This has been shown both for the ${\mathcal O}(p^4)$ and
${\mathcal O}(p^6)$ analytical formulae and checked numerically
as well. The latter can also be seen from
the fact that the contour plots below do not show a peaked
behavior near the diagonal where $\chi_4=\chi_1$.   

We have checked explicitly that the self-energy when calculated for
the different channels has the structure of 
Eqs.~(\ref{Sigma_direct_product_structure})
and (\ref{Sigma_internal_structure}) as well as the relation of $\sigma$
to the charged case of Ref.~\cite{BDL}.

\subsection{The Numerics}
\label{numerics}

All plots in this section are given in terms of the relative shifts to the
lowest order
contribution ${\mathcal D}^{(2)}$. That is, we plot
\be
\Delta {\cal D}= \frac{\mathcal D}{{\mathcal D}^{(2)}}-1\,,
\ee
where we keep the contributions in Eq.~(\ref{defDNNLO})
up to and including ${\mathcal O}(p^4)$
when we refer to the NLO case and up to and including ${\mathcal O}(p^6)$
when we refer to the NNLO case.

In the long run, the input parameters should be determined by fits 
of the PQ$\chi$PT formulas to
lattice QCD data. These fits are not available at present,
although suitable simulation results should become available 
in the near future. 
So we present here
plots with the same choices of input parameters as in
\cite{BDL,BL1,BL2,BDL2}. This is mainly from
the NNLO fit to experimental data, referred to as ``Fit~10'', 
which is presented in Ref.~\cite{ABT2}. That fit has $F_0=87.7$~MeV and 
a renormalization scale of $\mu=770$ MeV. The NNLO LECs $K_i^r$ and 
the NLO LECs $L_4^r, L_6^r$ and $L_0^r$ were not determined in that fit
and they have been set to zero for simplicity.
It should be noted that $L_0^r$ cannot be determined from experimental data, 
since it is a distinguishable quantity only in the PQ theory. Some 
recent results on $L_4^r$ and $L_6^r$ have been obtained in 
Ref.~\cite{piK}, but they have nevertheless been set to zero in
the plots in this paper, since the present numerics are mainly 
intended for illustrative purposes.

\begin{figure}
\begin{center}
\includegraphics[angle=0,width=\columnwidth]{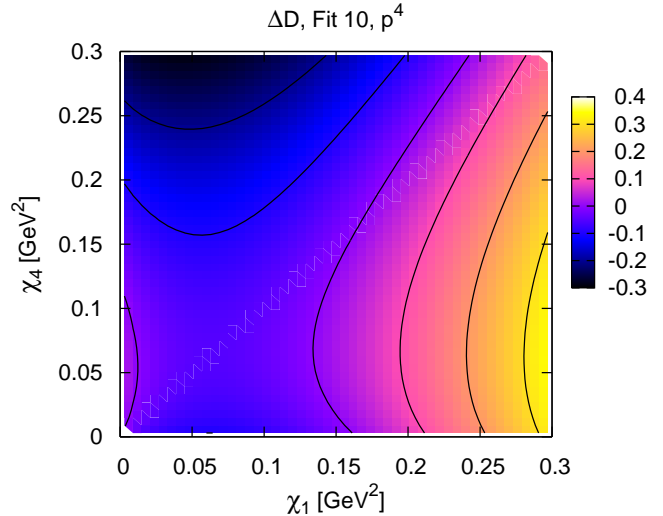}
\caption{The NLO relative shift of the lowest order contribution to the 
double pole coefficient $\cal D$, $\Delta{\cal D}$,
as a function of the valence and 
sea-quark masses $\chi_1$ and $\chi_4$. The difference between two 
successive contour lines in the plots is~$0.10$ and the values chosen for 
the LECs correspond to ``Fit~10'', as discussed in the text.}
\label{p4contour}
\end{center}
\end{figure}

\begin{figure}
\begin{center}
\includegraphics[angle=0,width=\columnwidth]{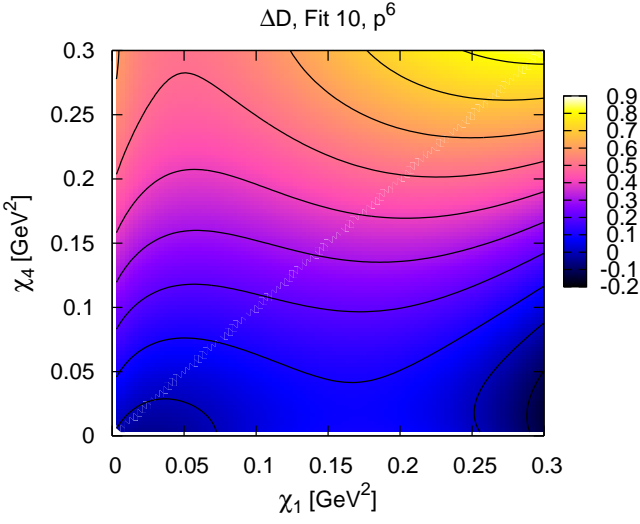}
\caption{The combined NLO and NNLO relative shifts of the lowest order 
contribution to the double pole coefficient $\cal D$, $\Delta{\cal D}$,
as a function 
of the valence and sea-quark masses $\chi_1$ and $\chi_4$. 
The difference between two successive contour lines in the 
plots is~$0.10$ and the values chosen for the LECs correspond to 
``Fit~10'', as discussed in the text.}
\label{p6contour}
\end{center}
\end{figure}

For the degenerate mass-case we are working with in this article, we have only
two input parameters for the quark masses, namely $\chi_1$ and $\chi_4$. This
allows for exhaustive contour plots of the whole parameter region of
interest. A contour plot of the NLO relative correction is given in
Fig.~\ref{p4contour}, and in Fig.~\ref{p6contour} the same parameter region is
plotted for the sum of the NLO and the NNLO shifts.
Let us remind the reader here that the quantities $\chi_i=2B_0 m_i$
correspond to lowest order meson masses squared. For the NNLO case,
diagrams appear which can have on-shell intermediate states for
$\chi_1>9\chi_4$. For that part of the plot we have simply kept the real
part only of the relevant sunset integral. The error due to this should be a
small effect,
similar in size to the width compared to the mass of the eta for realistic
quark masses.

The plots show a reasonable
convergence behavior for when both $\chi_1,\chi_4\rightarrow 0$. Furthermore,
as mentioned above, since we have plotted the relative correction with respect
to the lowest order contribution ${\mathcal D}^{(2)}=\chi_1-\chi_4$, the
plots would have a peaked limiting behavior near the diagonal line
$\chi_1=\chi_4$ if the NLO or NNLO corrections were nonzero for
$\chi_1=\chi_4$. That this is not the case is clear from the plots, and in
agreement with the expected result. Furthermore, the general shape of the
contour over the plotted parameter space is rather different for the NLO and
the NNLO results, in particular, the ${\mathcal O}(p^6)$ contributions
are rather sizable. This is an expected result as well, since previous
calculations of the masses and decay constants to NNLO show exactly this
behavior. In Ref.~\cite{BDL2} it was shown that the type of behavior
is strongly dependent on the values of the largely unknown NNLO LECs
and an example of changes in some of the less
well determined LECs was given that improves the convergence behavior
dramatically.

However, since a major point of this article is the dependence on
the LEC $L_7^r$ for ${\mathcal D}$, we have, in addition to the contour plots,
chosen a parameter line in the
two-dimensional mass-space, where results for different values of
the LECs have been plotted. This parameter line is defined by
\be
\chi_4 = \tan 60^\circ\,\chi_1,
\label{thxfact}
\ee
This choice is motivated by the fact that the sea quark mass is usually taken
to be larger than the valence quark mass in lattice simulations.

\begin{figure}
\begin{center}
\includegraphics[width=\columnwidth]{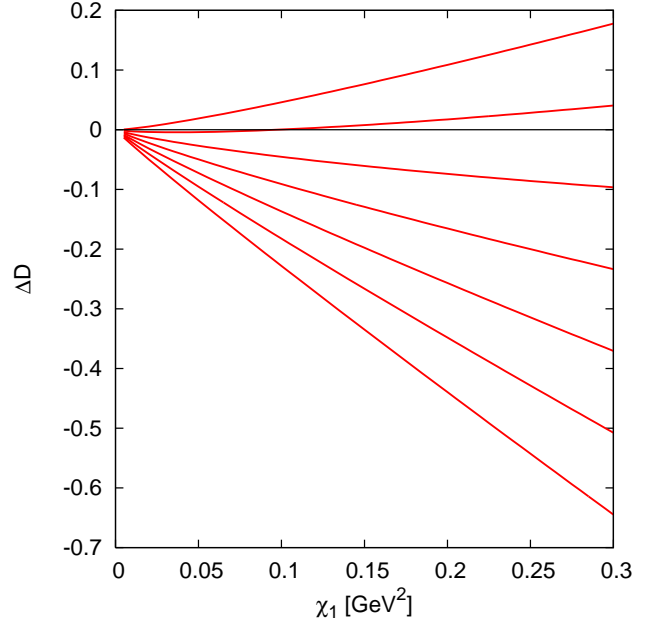}
\caption{The NLO relative shift of the lowest order 
contribution to the double pole coefficient $\cal D$, $\Delta{\cal D}$,
plotted
for $\theta=60^\circ$ in the $\chi_1 - \chi_4$ 
plane, for $10^4\,L_7^r=\{0, -1,-2,-3,-4,-5,-6\}$ (top to bottom curve).
The values chosen for the other LECs correspond to ``Fit~10'' as discussed
in the text.}
\label{p4ray}
\end{center}
\end{figure}

In Fig.~\ref{p4ray} the NLO relative correction has been plotted along this
parameter line, as a function of $\chi_1$. The plot is for several different
values of the LEC $L_7^r$, ranging between zero and $-0.0006$. All other
parameters are set to the values of ``Fit~10''. 
Convergence as $\chi_1\rightarrow 0$, for all values of $L_7^r$, is
evident, in agreement with the contour plot. 
The dependence on $L_7^r$ is exactly linear as expected from \cite{Sharpe2}
and Eq.~(\ref{resultp4}).

\begin{figure}
\begin{center}
\includegraphics[width=\columnwidth]{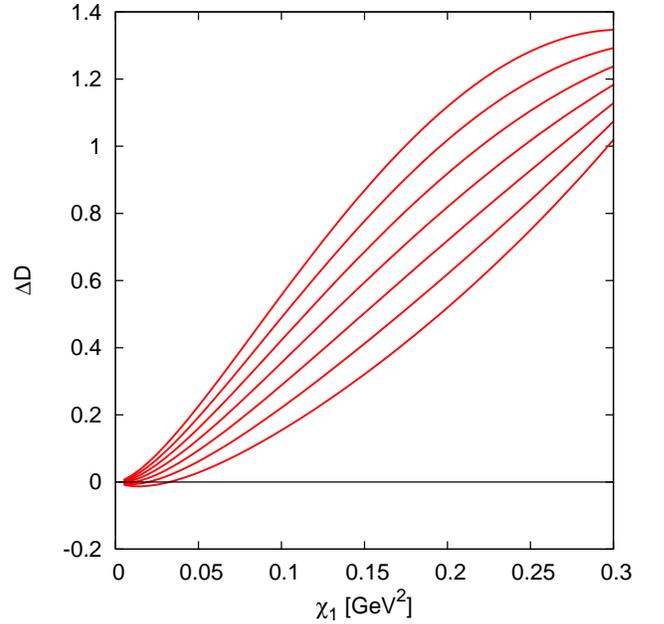}
\caption{The NNLO relative shift to the lowest order 
contribution to the double pole coefficient $\cal D$, $\Delta{\cal D}$,
plotted for $\theta=60^\circ$ in the $\chi_1 - \chi_4$ 
plane, for $10^4\,L_7^r=\{0,-1,-2,-3,-4,-5,-6\}$ (top to bottom curve).
The values chosen for the other LECs correspond to ``Fit~10''.}
\label{p6ray}
\end{center}
\end{figure}

\begin{figure}
\begin{center}
\includegraphics[width=\columnwidth]{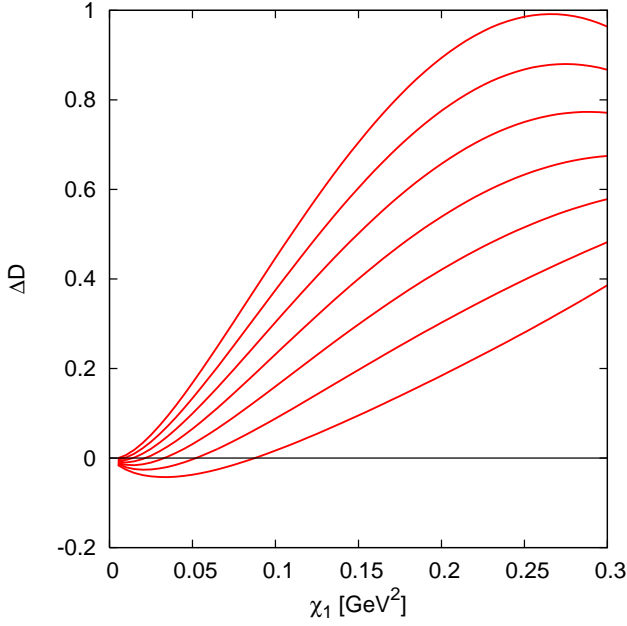}
\caption{The NNLO relative shift to the lowest order 
contribution to the double pole coefficient $\cal D$, $\Delta{\cal D}$,
plotted for $\theta=60^\circ$ in the $\chi_1 - \chi_4$ 
plane, for $10^4\,L_7^r=\{0, -1, 
-2, -3,-4,-5,-6\}$ (top to bottom curve). 
The values chosen for the other LECs are all set to zero.}
\label{p6rayNoLEC}
\end{center}
\end{figure}

\begin{figure}
\begin{center}
\includegraphics[width=\columnwidth]{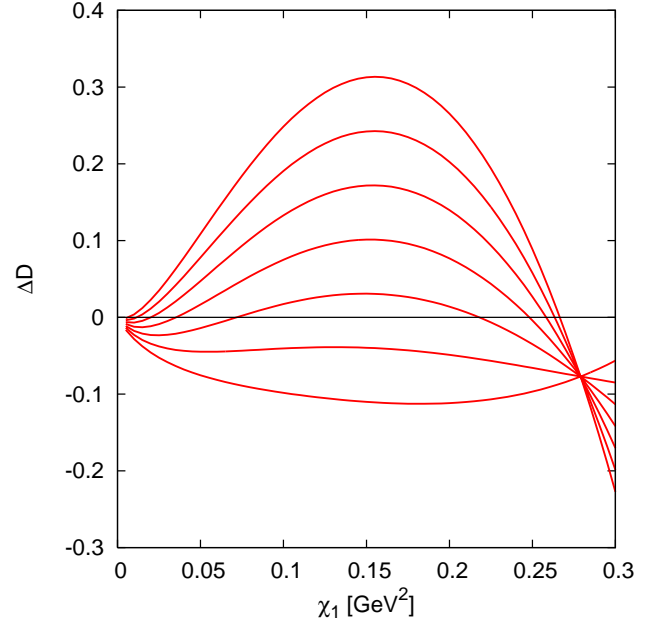}
\caption{The NNLO relative shift to the lowest order 
contribution to the double pole coefficient $\cal D$, $\Delta{\cal D}$,
plotted for $\theta=60^\circ$ in the $\chi_1 - \chi_4$ 
plane, for $10^4\,L_7^r=\{0, -1, 
-2, -3,-4,-5,-6\}$ (top to bottom curve). 
The values chosen for the other LECs are all set to zero.
The subtraction scale is $\mu=0.6$~GeV.}
\label{p6rayNoLECmu}
\end{center}
\end{figure}

In Fig.~\ref{p6ray}, the NNLO relative correction is plotted using the same
values for the LECs, namely ``fit~10'', except for $L_7^r$. As a comparison,
 Fig.~\ref{p6rayNoLEC} shows the NNLO relative correction with all LECs,
 except $L_7^r$, set to zero. As anticipated from the contour plot, the
contribution is sizable in both cases, but with a somewhat lower overall
amplitude when the LECs are set to zero. Also for the NNLO results, we have an
exact linear dependence on $L_7^r$ as can be seen from
Eq.~(\ref{resultp6loops}). As already discussed, the large
amplitude of the NNLO corrections is not expected to be a concern for the
convergence of the theory, since many of the LECs are as of yet poorly
determined. As an example to show this as well as the subtraction
scale dependence,
we have shown in Fig.~\ref{p6rayNoLECmu} the NNLO relative
correction as a function of $L_7^r$ and all other LECs set to zero
at a subtraction scale $\mu=0.6$~GeV.

\section{Fitting strategies}
\label{fitting_strategies}

\subsection{LECs relevant for the eta mass}

The eta mass to NNLO in $\chi$PT was calculated in \cite{ABT2}
and to NLO in \cite{GL2}.
We present here the corrections in the form
with the dependence written in terms of lowest order masses
and decay constants as calculated in \cite{ABT2} and obtainable from
\cite{website}. The NLO part depending on the $p^4$ LECs
is
\ba
m_\eta^{2(4)}&\!\sim\!&
 \hat m m_s   (  - 40/3 L_4^r - 32/9 L_5^r + 80/3 L_6^r - 64/3 L_7^r )
\nonumber\\&&
 + \hat m^2   (  - 16/3 L_4^r - 8/9 L_5^r + 32/3 L_6^r + 32/3 L_7^r
\nonumber\\&&
 + 16/3 L_8^r )
       + m_s^2   (  - 16/3 L_4^r - 32/9 L_5^r + 32/3 L_6^r
\nonumber\\&&
 + 32/3 L_7^r
 + 32/3 L_8^r )\,.
\ea
The $L_i^r$ occur in the three combinations
$2L_6^r-L_4^r$, $2L_8^r-L_5^r$ and $L_5^r+6L_7^r$.

The NNLO part depending on the NNLO LECs is
\ba
m_\eta^{2(6)}&\!\!\sim\!\!&
        \hat m m_s^2   (  - 128/9 C^r_{12} - 128/3 C^r_{13} - 32/9 C^r_{14}
\nonumber\\&&
 - 64/3 C^r_{15} - 16/
         3 C^r_{16} - 32/9 C^r_{17} + 32/3 C^r_{18}
\nonumber\\&&
 + 48 C^r_{20}
 + 144 C^r_{21} + 128/3 C^r_{32} - 64/
         3 C^r_{33} )
 \nonumber\\&&
      + \hat m^2 m_s   (  - 64/9 C^r_{12} - 32 C^r_{13} - 32/9 C^r_{14}
\nonumber\\&&
 - 16 C^r_{15} - 64/3 
         C^r_{16} - 32/9 C^r_{17} + 32 C^r_{20} + 192 C^r_{21}
\nonumber\\&&
 + 32/3 C^r_{32}
 - 64/3 C^r_{33} )
       + \hat m^3   (  - 32/27 C^r_{12}
\nonumber\\&&
 - 64/9 C^r_{13} - 16/9 C^r_{14}
 - 32/9 C^r_{15} - 32/3 
         C^r_{16}
\nonumber\\&&
 - 16/9 C^r_{17} - 32/9 C^r_{18} + 16 C^r_{19} + 32 C^r_{20}
 + 64 C^r_{21} 
\nonumber\\&&
+ 32/3 
         C^r_{31} + 64/3 C^r_{32} + 64/3 C^r_{33} )
\nonumber\\&&
       + m_s^3   (  - 256/27 C^r_{12} - 128/9 C^r_{13} - 64/9 C^r_{14}
 \nonumber\\&&
- 64/9 C^r_{15} - 32/3 
         C^r_{16} - 64/9 C^r_{17} - 64/9 C^r_{18}
\nonumber\\&&
 + 32 C^r_{19} + 32 C^r_{20}
 + 32 C^r_{21} + 64/3 
         C^r_{31} + 64/3 C^r_{32}
\nonumber\\&&
 + 64/3 C^r_{33} )\,.
\ea

The $C_i^r$ contributions there can be rewritten in terms of the three
flavor PQ$\chi$PT NNLO LECs $K_i^r$ using the Cayley-Hamilton relations
of \cite{BCE1}. The arguments for this were presented in detail in \cite{BDL2}.
The resulting formula in terms of the combinations of $K_i^r$ defined below
is
\ba
m_\eta^{2(6)}&\!\!\sim\!\!&
       \hat m m_s^2   (  - 32/9 X_2 - 128/3 X_3 - 48 X_4 + 128/3 X_5
\nonumber\\&&
 - 64/9 X_6 )
       + \hat m^2 m_s   (  - 32/9 X_2 - 32 X_3 - 64 X_4
\nonumber\\&&
 + 128/3 X_5 - 64/9 X_6 + 64/9 
         X_7 )
       + \hat m^3   (  - 32/3 X_1
\nonumber\\&&
 + 80/9 X_2 - 64/9 X_3 - 64/3 X_4 + 32/3 X_5
\nonumber\\&&
 + 64/9 
         X_6 - 128/27 X_7 )
       + m_s^3   (  - 64/3 X_1 
\nonumber\\&&
+ 128/9 X_2 - 128/9 X_3 - 32/3 X_4 + 64/9 X_6
\nonumber\\&&
 - 64/
         27 X_7 )\,.
\ea

\subsection{Obtaining LECs from $\D$: fitting strategies}

In Ref.~\cite{BDL2}
 we discussed at some length, possible fitting strategies for
determining the various LECs at ${\mathcal O}(p^4)$ and ${\mathcal O}(p^6)$.
In the analytical expressions for ${\mathcal D}$ some additional
LECs are present, which motivates further elaboration on these fitting
strategies. At $O(p^4)$ the only new LEC is $L_7^r$ and at
${\mathcal O}(p^6)$ we now have $K_{24}^r, K_{41}^r$ and $K_{42}^r$ as well. 
It turns out
that  together with the expressions for ${\mathcal D}$ it is
possible to determine all the combinations of LECs which appear in the
NLO+NNLO expressions for the masses including the eta mass.
We will now give a short outline of how one might
proceed but we will concentrate on the masses only.

From the charged masses we have the dependence 
\begin{eqnarray}
\delta^{(4)23}_{\mathrm{ct}} & \sim &
3\bar\chi \left(2L_6^r-L_4^r\right) \:+\:
\chi_{13} \left(2L_8^r-L_5^r\right)\,.
\label{massL}
\end{eqnarray} 
In these expressions, $\bar \chi$ denotes the average over the sea-sector
quarks, and the superscripts $23$ indicate having 2
nondegenerate valence quark masses and 3 nondegenerate sea quark
masses. We
refer to the discussion in Ref.~\cite{BDL2} for details on which mass
degeneracies
which are needed for the determination of the LECs, and just refer to the
needed expressions here.

The NLO result for ${\mathcal D}$ is proportional to 
\be
{\mathcal D^{(4)}_{\mathrm{ct}}}\sim 3 \chi_4 (2L_6^r-L_4^r)
+2\chi_{14}(2L_8^r-L_5^r)-R_{14}^d (L_5^r+6L_7^r)
\ee
from which an overall factor of $R_{14}^d$ has been removed. 
The resemblance
to the mass expressions is striking, even though the coefficients are not
quite the same.
However, it is clear that all combinations needed to determine the
eta mass at NLO can now be obtained.
Since we can obtain $L_5^r$ from the decay constant we can also get
at $L_7^r$ separately. This was already shown to NLO in \cite{Sharpe1,Sharpe2}.

Let us now do the same thing for the NNLO LECs.
The situation is somewhat more involved but not different in principle.
The full discussion can be found in \cite{BDL2}, we only present
a part of it here. The mass at NNLO for the (2+2) case there
depended on the NNLO LECs as
\ba
\label{massKi}
\delta^{(6)22}_{\mathrm{ct}} &\!\!=\!\!&
        \chi_1 \chi_3 \chi_4   (  - 32 X_3 )
       + \chi_1 \chi_3 \chi_6   (  - 16 X_3 )
\nonumber\\&&
       + \chi_1 \chi_3^2   (  - 12 X_1 + 8 X_2 )
       + \chi_1 \chi_4 \chi_6   (  - 32 X_4 + 32 X_5 )
\nonumber\\&&
       + \chi_1 \chi_4^2   (  - 32 X_4 + 16 X_5 )
       + \chi_1 \chi_6^2   (  - 8 X_4 )
\nonumber\\&&
       + \chi_1^2 \chi_3   (  - 12 X_1 + 8 X_2 )
       + \chi_1^2 \chi_4   (  - 16 X_3 )
\nonumber\\&&
       + \chi_1^2 \chi_6   (  - 8 X_3 )
       + \chi_1^3   (  - 4 X_1 )
\nonumber\\&&
       + \chi_3 \chi_4 \chi_6 32  (  -  X_4 +  X_5 )
       + \chi_3 \chi_4^2   (  - 32 X_4 + 16 X_5 )
\nonumber\\&&
       + \chi_3 \chi_6^2   (  - 8 X_4 )
       + \chi_3^2 \chi_4   (  - 16 X_3 )
       + \chi_3^2 \chi_6   (  - 8 X_3 )
\nonumber\\&&
       + \chi_3^3   (  - 4 X_1 )\,.
\ea
The combinations $X_i$ are defined as
\ba
X_1 &=& -K^r_{39}+K^r_{17}+K^r_{19}-3 K^r_{25}  \nonumber\\
X_2 &=& K^r_{19}-K^r_{23}-3 K^r_{25}  \nonumber\\
X_3 &=& K^r_{18}+K^r_{20}/2-K^r_{26}-K^r_{40}  \nonumber\\
X_4 &=& K^r_{22}+K^r_{21}-K^r_{26}-3 K^r_{27}  \nonumber\\
X_5 &=& K^r_{21}-K^r_{26}  \nonumber\\
X_6 &=& K^r_{23}+2 K^r_{17}-3/2 K^r_{24}+3/2 K^r_{25}+3 K^r_{41 } \nonumber\\
X_7 &=& K^r_{17}-3/2 K^r_{24}-3 K^r_{26}-3 K^r_{40}-9 K^r_{42}\,.
\ea
From Eq.~(\ref{massKi}) it is clear that the combinations $X_1$ to $X_5$
can be determined from the charged masses at NNLO by varying the input quark
masses in the lattice QCD simulations.
The above is a rephrasing of the discussion of \cite{BDL2}.

Let us now rewrite the NNLO LECs contribution to $\cal D$ in the same
fashion. Eq.~(\ref{resultp6Ki}) can be rewritten as
\ba
{\mathcal D}^{(6)}_{\mathrm{ct}} & \sim &
        \chi_1 \chi_4  (  - 2 X_1 + X_2 - 12 X_3 + 2 X_7 )
       + \chi_1^2  (  - 2 X_1
\nonumber\\&&
 - 2 X_6 )
       + \chi_4^2  (  - 2 X_1 + 2 X_2 - 9 X_4 + 6 X_5 + 2 X_6
\nonumber\\&&
 - 2 X_7 )\,,
\ea
It is clear that from $\cal D$ we can thus determine the remaining LECs needed
to get at the eta mass at NNLO. We can easily get at the missing combinations
$X_6$ and $X_7$.

\section{Conclusions}
\label{conclusions}

In this paper, we have calculated the quantity ${\mathcal D}$ to NNLO. This
quantity was first proposed and calculated to NLO by Sharpe and
Shoresh~\cite{Sharpe1,Sharpe2},
as a way to obtain a better fit of the LEC $L_7^r$, since it
appears already at NLO for this quantity. This then also allowed to get the
eta mass at NLO in $\chi$PT.
As discussed in~\cite{Sharpe1}, ${\mathcal  D}$
is a well defined Lattice quantity in terms of the correlation
function $R_0(t)$.

We have presented the full analytical results for ${\mathcal D}$ to NNLO,
together with a
numerical analysis indicating the typical size of the corrections for various
choices of the LECs. In particular, we demonstrate the sensitive dependence on
the LEC $L_7^r$ which these expressions exhibit, both at NLO and at NNLO. It
should also be noted that, in addition to showing dependence on the LEC
$L_7^r$ already at NLO,
the NNLO analytical expressions presented here also depend on the LECs
$K_{24}^r, K_{41}^r$ and $K_{42}^r$, none of which were present in the NNLO
expressions for the charged masses or decay constants. 

We have extended the discussion about possible fitting strategies
in Ref.\cite{BDL2} to include the new LECs which appear in the NNLO results for
${\mathcal D}$. The conclusion is that the NLO and NNLO expressions for the
charged masses and decay constants, together with the expressions for
${\mathcal D}$ provided in this article, in fact give enough information to
determine
all of the combinations of LECs which appear in these calculations,
including those for the eta mass at NNLO.

In addition we have also shown how to resum explicitly the full propagator
from the lowest order propagator and self-energies also in the case for the
neutral propagator in PQ$\chi$PT.

\section*{Acknowledgments}

The program \texttt{FORM 3.0} has been used extensively in these 
calculations~\cite{FORM}. This work is supported by the European Union 
TMR network, Contract No. HPRN - CT - 2002 - 00311 (EURIDICE) and the 
EU - Research Infrastructure Activity RII3 - CT - 2004 - 506078 
(HadronPhysics). 

\appendix

\section{Symmetry Constraints on the propagator $G_{ijkl}$}
\label{Appmatrix}

In Ref.~\cite{Sharpe2}, App. E,
a number of constraints on the form of the meson
propagators
in PQQCD were derived. The Lagrangian of QCD
played no role, since the only properties of the quark bilinears which were
used were the transformation properties under vector symmetries, whose
properties are shared by the meson fields of PQ$\chi$PT. It was then argued
that the derivation presented in~\cite{Sharpe2} could easily be extended to
the case
with more than one flavor of valence quarks and nondegenerate masses.
We will now show this
generalization to the general case with $n_\mathrm{nval}$ valence and ghost
quarks and $n_\mathrm{sea}$ sea quarks.
The masses are fully general except that the $a$-th ghost quark has the
same mass as the $a$-th valence quark mass for all
$a=1,\ldots,n_\mathrm{nval}$. 
Three types of symmetries are needed, all fairly trivial extensions
of the ones in \cite{Sharpe2}.

We suppress the time ordering and often the space-time dependence
in the remainder of this appendix to make the symmetry structures more visible.

\subsection{Phase rotations and $G^{nT}=G^n$}

We write the quark
flavors as a vector
\be
\mathbf q=(q_1,\dots,q_{2n_\mathrm{nval}+n_\mathrm{nsea}})^T,
\ee 
where indices $1$ to $n_\mathrm{nval}$ denote the valence quarks.
Indices
$n_\mathrm{nval}+1$ to $n_\mathrm{nval}+n_\mathrm{nsea}$ denote the sea quarks
and indices
$n_\mathrm{nval}+n_\mathrm{nsea}+1$
to $2n_\mathrm{nval}+n_\mathrm{nsea}$ denote the bosonic ``ghost'' quarks.

The direct generalization of the subgroup $V$
of the vector transformations in Ref.~\cite{Sharpe2}, is
\be
V=\diag\left(\exp \theta_1,\dots,
\exp \theta_{2n_\mathrm{nval}+n_\mathrm{nsea}}\right).
\ee
Under a phase rotation of only the flavor $m$, $G_{ijkl}$ transforms as
\be
G_{ijkl}\rightarrow G_{ijkl}\exp\left(\theta_m(\delta_{im}-\delta_{jm}+\delta_{km}-\delta_{lm})\right)\,.
\ee
The invariance of $G_{ijkl}$ 
implies that the indices must be paired up. Thus, non-vanishing elements of
$G$ have the form $G_{iijj}$ or $G_{ijji}$.

A second relation which can be simply derived is that $G^n$ is
a symmetric matrix.
Consider
\begin{eqnarray}
\langle \Phi_{ij}(x)\Phi_{kl}(0)\rangle &=& \langle
\Phi_{ij}(-x)\Phi_{kl}(0)\rangle\nonumber\\
&=& \langle \Phi_{ij}(0)\Phi_{kl}(x)\rangle\nonumber\\
&=& (-)\langle \Phi_{kl}(x)\Phi_{ij}(0)\rangle,
\end{eqnarray}
which make use of rotation invariance and translation invariance. The minus
sign in the last step is only needed when both $\Phi_{ij}$ and
$\Phi_{kl}$ are fermionic fields. This equation implies that
$G_{ijkl}=(-)G_{klij}$. In particular, this implies that $G^n$ defined
in (\ref{structureneutral}) is a symmetric matrix.

At this level $G^n$ has the structure
\be
\label{structureG1}
G^n = \left(\begin{array}{ccc}r+s&t^T&v^T\\t&u&w\\v&w^T&r-s
  \end{array}\right)\,.
\ee
$r,s,u$ are symmetric matrices. The matrices are ordered in the valence,
sea and bosonic sectors.

\subsection{$SU(1|1)$ transformations}

The third type of symmetries is $SU(1|1)$ transformations involving 
the $m$-th
quark from the valence sector $q_v$ and the $m$-th
 quark from the bosonic sector $q_b$. There are $n_\mathrm{val}$
of such $SU(1|1)$ symmetries. Below we will use the label $m_V$
and $m_B$ to denote these quarks respectively. 

This type of transformations is
of the form
\be
\label{SU11}
U=\left(\begin{array}{cc} a & b \\ c & d \end{array}\right)\in
\mathrm{SU}(1|1)_m .
\ee
where $a,d$ are commuting numbers, 
$b,c$ are anticommuting numbers and $U$
satisfies $UU^\dagger=\mathbf{I}_{2\times 2}$. Here we think of it
as acting on the vector
\be
\mathbf Q=(q_{m_V},q_{m_B})^T\,.
\ee 
The index $m$ in (\ref{SU11}) indicates which $SU(1|1)$ we are referring to.

This type of symmetry transformations can, together with the previous
symmetries,
be used to set further constraints on the block structure of $G_{ijkl}$.
In order to see this, we
first form the following classes of 2-indexed objects out of elements of $G$:
\begin{eqnarray}
O^1_{ik} &=&\sum_{j=m_V,m_B}G_{ijjk}(x)  
\nonumber\\
O^2_{ik} &=&\sum_{j=m_V,m_B} \varepsilon_j G_{jjik}(x)
\nonumber\\
O^3_{ik} &=& G_{n_Sn_Sik}(x) 
\nonumber\\
O^4_{ik} &=&G_{in_Sn_Sk}(x)\,
\label{symmetries1}
\end{eqnarray}
where  $n_S$ indicates the $n$-th index in the sea sector, and the matrix
indices
$i,k$ take both take values from $\{m_V,m_B\}$
so as to
form a supermatrix (or graded) structure. 
The symbol $\varepsilon_j$ gives the correct sign of the graded symmetry,
i.e. $\varepsilon_j\equiv 1$ for $j=1,\dots,n_\mathrm{val}+n_\mathrm{sea}$
and $\varepsilon_j\equiv -1$ for the remaining ones.
The objects in(~\ref{symmetries1}) are the direct generalizations of the
objects in~\cite{Sharpe2} to the
general flavor case. 

In addition to these, we will also need the two classes of objects
defined by
\begin{eqnarray}
O^5_{ik} &=&G_{n_Vn_Vik}(x)\,,
\nonumber\\
O^6_{ik} &=&G_{n_Bn_Bik}(x)\,.
\label{symmetries2}
\end{eqnarray}
Here, the value chosen for $n_V$ must be different from the value chosen for
$m_V$ and $n_B$ different from $m_B$. The subscripts $V$ and $B$ indicate
that $n_V$ is a valence index and $n_B$ a ghost index.
All the objects defined above transform under $SU(1|1)_m$ as
\be
O^l_{ik}\rightarrow \sum_{o,p=m_V,m_B}U_{io}O^l_{op}U_{pk}^{\dagger}.
\label{transformation}
\ee
Since $SU(1|1)_m$ is a symmetry, the expressions are invariant under these
transformations. 
Writing out these transformations explicitly one arrives, after some
calculations, at a number of constraint equations, which must be satisfied,
for the symmetry to hold. Solving these equations, the constraints following
from~(\ref{symmetries1}) are
\begin{eqnarray}
G_{m_Vm_Vm_Vm_V} &=& r_{mm}+s_{mm}\nonumber\\
G_{m_Bm_Bm_Bm_B} &=& r_{mm}-s_{mm}\nonumber\\
G_{m_Vm_Vm_Bm_B} &=& G_{m_Bm_Bm_Vm_V} = r_{mm}\nonumber\\
G_{m_Bm_Vm_Vm_B} &=& -G_{m_Vm_Bm_Bm_V} =s_{mm}\nonumber\\
G_{n_Sn_Sm_Vm_V} &=& G_{m_Vm_Vn_Sn_S} = G_{n_Sn_Sm_Bm_B}
\nonumber\\
 &=& G_{m_Bm_Bn_Sn_S} =t_{mn}\nonumber\\
G_{m_Vn_Sn_Sm_V}&=& G_{n_Sm_Vm_Vn_S} = G_{m_Bn_Sn_Sm_B}
\nonumber\\
& =&
-G_{n_Sm_Bm_Bn_S}\,,
\label{Gsym1}
\end{eqnarray}
where as defined above, $m_V$ is the $m$-th valence quark index
$m_B$ is the $m$-th ghost quark index and $n_S$ is the $n$-th sea quark index.
What we have shown so far is that the matrices
$t$ and $w$ in (\ref{structureG1}) are identical
and that the diagonal parts of $r$ and $v$ are the same.

The previous was a simple extension of what was done in \cite{Sharpe2}.
The really new part is that for the general flavor
case, we also have a
distinction between diagonal elements and non-diagonal elements in the
various matrices of $G^n$. The invariant objects
in~(\ref{symmetries2}) provides information about the matrix structure of
these. Writing out the
transformation~(\ref{transformation}) for these objects yields the additional
constraint equations
\begin{eqnarray}
G_{n_Vn_Vm_Vm_V} &=& G_{n_Vn_Vm_Bm_B} = r_{nm}\nonumber\\
G_{n_Bn_Bm_Vm_V} &=& G_{n_Bn_Bm_Bm_B},
\label{Gsym2}
\end{eqnarray}
for $n_V \ne m_V, n_B \ne m_B$.
These equations imply that the off-diagonal part of $s$ vanishes
and that the off-diagonal parts of $v$ and $r$ are the same.

To summarize,
$G^n$ has the structure
\be
\label{structureG2}
G^n = \left(\begin{array}{ccc}r+s&t^T&r\\t&u&t\\r&t^T&r-s
  \end{array}\right)\,.
\ee
$r,s,u$ are symmetric matrices and $s$ is diagonal.

Here we can see that since $t$ and $r$ connect different sectors
they are disconnected if we think in terms of external quark lines.
That $s$ is from connected lines and related to charged propagators
is a little more involved to show.
First, we use the fourth line
in Eq.~(\ref{Gsym1}) where we have an obviously charged
object in the propagator. We now set a second valence/ghost quark
to the same mass, i.e. $m_{m_V}=m_{n_V}$. We can then use the permutation
symmetry to turn the $m_V$ indices into $n_V$.
The final step is to use the $SU(1|1)_m$ symmetry to rotate the $m_B$
index back to $m_V$.
\be
s_{mm} = G_{m_B m_V m_V m_B}
= G_{m_B n_V n_V m_B}
= G_{m_V n_V n_V m_V},
\ee
which is the desired relation between $s$ and the charged meson
propagator.

\subsection{Equal mass limits}

When masses are equal, we have more symmetries.

For all valence quark masses equal we can interchange two
indices $m_V$
and $n_V$ everywhere and $G^n$ should remain invariant.
This implies that we can interchange columns in $t$ and $v$ of
Eq.~(\ref{structureG1}).
It also implies that $r+s$ must be invariant under the simultaneous
change of a pair of columns and rows.

For all bosonic quark masses equal we can interchange two indices $m_B$
and $n_B$ everywhere and $G^n$ should remain invariant.
This implies that we can interchange rows in $v$ and $w^T$ of
Eq.~(\ref{structureG1}).
It also implies that $r-s$ must be invariant under the simultaneous
change of a pair of columns and rows.

For all sea quark masses equal we can interchange two indices $m_S$
and $n_S$.
This means that $t$ and $w$ are invariant under the interchange
of rows and that $u$ is invariant under the simultaneous interchange
of rows and columns.

Putting the above together with $t=w$ and $r=v$ derived earlier
shows that $r$ and $t$ are matrices with all elements equal.
$s$ was diagonal so all its diagonal elements must be equal.
For $u$ it implies that all diagonal elements are equal
and that all off-diagonal elements are equal. The matrix then has
precisely the structure of Eq.~(\ref{G_blocks}).

\section{The resummation in the general case}
\label{Appresum}

In this appendix we show how to do the resummation in the general case
for an arbitrary number of valence quarks $n_\mathrm{val}$
and sea quarks $n_\mathrm{sea}$.

The general lowest order neutral propagator matrix (\ref{propn})
can be written in the form
\be
\label{LOpropfull}
G^0 = i 
\left(\begin{array}{ccc}\alpha&0&0\\0&\beta&0\\0&0&-\alpha\end{array}\right)
-i\delta
\left(\begin{array}{c}\tilde\alpha\\\tilde\beta\\\tilde\alpha\end{array}\right)
\left(\begin{array}{ccc}
\tilde\alpha^T&\tilde\beta^T&\tilde\alpha^T\end{array}\right)\,.
\ee
The notation here is similar but not the same as in the main text.
The matrices $\alpha$ and $\beta$ are diagonal matrices with propagators.
\ba
\alpha &=& \mathrm{diag}\left(\alpha_1,\ldots,\alpha_{n_\mathrm{val}}\right)\,,
\nonumber\\
\beta &=& \mathrm{diag}\left(\beta_1,\ldots,\beta_{n_\mathrm{sea}}\right)\,,
\nonumber\\
\tilde\alpha^T &=& \left(\alpha_1,\ldots,\alpha_{n_\mathrm{val}}\right)\,,
\nonumber\\
\tilde\beta^T &=& \left(\beta_1,\ldots,\beta_{n_\mathrm{sea}}\right)\,,
\nonumber\\
\alpha_i &=& 1/\left(p^2-\chi_i\right)\,,
\nonumber\\
\beta_i &=& 1/\left(p^2-\chi_{n_\mathrm{val}+i}\right)\,,
\nonumber\\
\delta &=& \frac{1}{\sum_{j=1,n_\mathrm{sea}}\beta_j},.
\ea
The the various $\alpha$ symbols are the valence and ghost sectors and $\beta$
the sea quark sector.
In the equal mass limit Eq.~(\ref{LOpropfull}) reduces to Eq.~(\ref{LOprop}).

The self-energy has the structure derived in App.~\ref{Appmatrix}
where the submatrices now have a more general structure than in the main text.
\be
\label{Sigmaapp}
\Sigma = \left(\begin{array}{ccc}
R+S&T^T&-R\\T&W&-T\\-R&-T^T&R-S\end{array}\right)\,.
\ee
$R$ is a symmetric $n_\mathrm{val}$ by $n_\mathrm{val}$ matrix.
$W$ is a symmetric $n_\mathrm{sea}$ by $n_\mathrm{sea}$ matrix.
$S$ is a diagonal $n_\mathrm{val}$ by $n_\mathrm{val}$ matrix.
$T$ is a $n_\mathrm{sea}$ by $n_\mathrm{val}$ matrix.
$R$ and $T$ have only disconnected, $S$ only connected parts when
thinking in terms of the external quark lines. $W$ has contributions
from both.

The full propagator in the neutral sector is the sum
\be
G = \sum_{n=0,\infty} \left(G^0(-i)\Sigma\right)^n G^0\,.
\ee
Not all parts in $\Sigma$ will appear in the final resummation to all
powers. We thus need to extract the parts that appear to all orders
similar to what was done in the main text by splitting into $\mathbf{I}$
and $\mathbf{K}$. We do it here by splitting $\left(G^0(-i)\Sigma\right)$
into four parts with different properties.
\be
G^0(-i)\Sigma = A+B+C+D\,,
\ee
with
\ba
A &\!=\!& 
\left(\begin{array}{ccc}\alpha S&0&0\\0&\beta W&0\\0&0&\alpha S\end{array}
\right)
\nonumber\\
B &\!=\!& 
\left(\begin{array}{ccc}\alpha R&0&-\alpha R\\0&0&0\\\alpha R&0&-\alpha R
\end{array}\right)
\nonumber\\
C &\!=\!& 
\left(\begin{array}{ccc}0&\alpha T^T&0\\\beta T&0&-\beta T\\0&\alpha T^T&0
\end{array}\right)
\nonumber\\
D &\!=\!&\!
\left(\begin{array}{c}\!-\delta\tilde\alpha\\\!-\delta\tilde\beta
\\\!-\delta\tilde\alpha\end{array}\right)\!
\left(\begin{array}{ccc}
\tilde\alpha^T S +\tilde\beta^T T&\tilde\beta^T W&
 -\tilde\alpha^T S-\tilde\beta^T T\end{array}\right)\,.
\nonumber\\
\ea
We define four general types of structures
\ba
\mathrm{I} &&\left(\begin{array}{ccc}X&0&0\\0&Y&0\\0&0&X\end{array}\right)\,,
\nonumber\\
\mathrm{II}&&
\left(\begin{array}{ccc}Z&0&-Z\\0&0&0\\Z&0&-Z\end{array}\right)\,,
\nonumber\\
\mathrm{III}&&
\left(\begin{array}{ccc}0&U&0\\V&0&-V\\0&U&0\end{array}\right)\,,
\nonumber\\
\mathrm{IV}&&
\left(\begin{array}{c}\tilde x\\\tilde y\\\tilde x\end{array}\right)
\left(\begin{array}{ccc}\tilde u^T & \tilde v^T -\tilde u^T
\end{array}\right)\,.
\ea
The matrices $U,V,X,Y,Z$ and column vectors $u,v,x,y$ stand here for a generic
form.
These various types satisfy matrix multiplication properties.
\begin{itemize}
\item 
A product of a type I matrix with another one of type I,II,III,IV
is of type I,II,III,IV respectively. This is true for multiplication
by type I from front or back.
\item
A product of a type II matrix with another one of type II,III,IV
is zero. This is true for multiplication by type II from front or back.
\item
A product of two type III matrices is of type II.
\item
A product of type IV with type I,III,IV is of type IV.
\end{itemize}

The first property allows to treat the $A$ part similarly to
the $\mathbf{I}$ in the main text. We resum it separately and define
\be
\bar A \equiv \sum_{n=0,\infty} A^n\,.
\ee
Due to the diagonal structure of $\alpha$ and $\beta$ this resummation can
be done with
\be
\bar A =
\left(\begin{array}{ccc}\bar\alpha & 0&0\\0&\bar\beta &0\\0&0&\bar\alpha
\end{array}\right)
=
\left(\begin{array}{ccc}\frac{1}{1-\alpha S} & 0&0\\0&\frac{1}{1-\beta W} &0\\
0&0&\frac{1}{1-\alpha S}
\end{array}\right)\,.
\ee

The second property means that $B$ can only appear once,
and together with the
third means that $C$ multiplied by itself or products of $A$
appears at most twice.
Putting then all strings of $A$ only together in $\bar A$ the sum can be
rewritten as
\ba
\label{partialsum}
\sum_{n=0,\infty}\hskip-1ex\left(-iG^0\Sigma\right)^n\hskip-1ex&=&
\sum_{n=1,\infty}\hskip-1ex
\left((\bar A +\bar A C \bar A)D\right)^n 
(\bar A +\bar A C \bar A)
\nonumber\\&&
\!\!+\bar A+\bar A B \bar A +\bar A C \bar A+\bar A C \bar A C \bar A\,.
\nonumber\\
\ea
The remaining summation in (\ref{partialsum}) can be done by noticing that
\be
\left((\bar A +\bar A C \bar A)D\right)^2 =
\left((\bar A +\bar A C \bar A)D\right)
\left(-\delta \tilde\beta^T W \bar\beta\tilde\beta\right)\,.
\ee
The quantity in the last bracket is a scalar and can be taken out of the
matrix products in the partial sum. The sum then can be done by a geometric
series in only the quantity in the last set of brackets.

The remaining part of the calculation is to perform all of the
matrix multiplications. Some properties that are used heavily
to bring the result into its final form
are
\be
\frac{1}{1-\beta W}\beta = \beta\frac{1}{1-W\beta}\,,\quad
\left(\frac{1}{1-\beta W}\right)^T = \frac{1}{1-W\beta}\,.
\ee
There are similar properties for the $\alpha$ and $S$ combinations.

The result can be written in a form reminiscent of
the lowest order (\ref{LOpropfull}).
\be
\label{propfull}
G = i 
\left(\begin{array}{ccc}r+s&t^T&r\\t&w&t\\r&t^T&r-s\end{array}\right)
-i\bar\delta
\left(\begin{array}{c}\tilde a\\\tilde b\\\tilde a\end{array}\right)
\left(\begin{array}{ccc}
\tilde a^T&\tilde b^T&\tilde a^T\end{array}\right)\,,
\ee
where the matrices $r,s,t,w$ have the same structure as the matrices
$R,S,T,U$ in (\ref{Sigmaapp}).
These are given in terms of the self-energy and lowest order propagator
parts as
\ba
\label{propfull2}
s&=&\bar\alpha\alpha\,,
\nonumber\\
w&=&\bar\beta\beta\,,
\nonumber\\
r&=&\bar\alpha \alpha R\bar\alpha \alpha+\bar\alpha \alpha
T^T\bar\beta\beta T \bar\alpha \alpha\,,
\nonumber\\
t&=&\bar\beta\beta T \bar\alpha \alpha\,,
\nonumber\\
\tilde a &=& \bar\alpha\tilde\alpha+\bar\alpha T^T\bar\beta\tilde\beta\,,
\nonumber\\
\tilde b &=& \bar\beta\tilde\beta\,,
\nonumber\\
\bar\delta&=&\frac{\delta}{1+\delta \tilde\beta^T W \bar\beta\tilde\beta}\,.
\ea
Eqs. (\ref{propfull}) and (\ref{propfull2}) are the main results of this
appendix.
Note that all lowest order poles have been shifted to the resummed poles
everywhere where they appear. We have also checked that the results of this
appendix agree with those in the main text by taking the various equal
mass limits.

\end{document}